\documentclass[12pt]{article}
\usepackage{amsmath}
\usepackage{amssymb,verbatim,color}
\usepackage{bm,lscape}
\usepackage{graphicx}
\usepackage{wasysym}
\usepackage{theorem}
\usepackage[scriptsize]{subfigure}

\usepackage[authoryear]{natbib}

\allowdisplaybreaks

\def\bSig\mathbf{\Sigma}

\newcommand{\E}{\mathbb{E}}

\newcommand{\Nc}{\mathcal{N}}

\newcommand{\Uc}{\mathcal{U}}

\newcommand{\eps}{\varepsilon}

\newcommand{\Uno}{\mathbf{1}}

\newcommand{\be}{\begin{equation}}
\newcommand{\ee}{\end{equation}}
\newcommand{\iid}{\stackrel{\mathrm{iid}}{\sim}}
\newcommand{\ind}{\stackrel{\mathrm{ind}}{\sim}}

\DeclareMathOperator{\Var}{Var}

\theoremstyle{plain}

\allowdisplaybreaks

\begin{document}
\title{Bayesian Nonparametric Modelling of Joint Gap Time Distributions for  Recurrent Event Data}

\author{Marta Tallarita\footnotemark[1], Maria De Iorio, Alessandra Guglielmi and James Malone-Lee\\
UCL, London (UK) and Politecnico di Milano (ITALY)}
\date{\today}
\maketitle
\footnotetext[1]{\emph{E-mail:} m.tallarita@ucl.ac.uk}

\begin{abstract}
We propose autoregressive Bayesian semi-parametric models for waiting times between recurrent events. The aim is two-fold: inference on the effect of possibly time-varying covariates on the gap times and clustering of individuals based on the time trajectory of the recurrent event.   Time-dependency between gap times is taken into account through the specification of an autoregressive component  for the random effects parameters influencing the response at different times. The order of the autoregression may be assumed unknown and object of inference and we consider two alternative approaches to perform model selection under this scenario. Covariates may be easily included in the regression framework and censoring and missing data are easily accounted for. As the proposed methodologies lies within the class of Dirichlet process mixtures, posterior inference can be performed through efficient MCMC algorithms. We illustrate the approach through simulations and medical applications involving recurrent hospitalizations of cancer patients and successive urinary tract infections.
\end{abstract}


\section{Introduction}

Recurrent event processes generate events repeatedly over time and recurrent event data arise in many applications, for example in medicine, science and technology. 
Typical examples include recurrent infections, asthma attacks, hospitalizations, product repairs, machine failures. 
In particular, in this work, we are interested in settings where recurrent event processes are available from a large number of individuals, but with a small number of occurrences for each subject.
Typically, the focus is in modeling the rate of occurrence, accounting for the variation within and between individuals. Moreover, in applications, it is often of interest to assess the relationship between event occurrence 
and potential explanatory factors. The two main statistical approaches to perform inference on recurrent event data are \citep[see][]{CooLaw07}: $(i)$ modelling the intensity function of the event counts  $\{ N(t), t\geq 0\}$, where $N(t)$ is the number of events between the time origin and time $t$; $(ii)$ modelling the whole sequence of waiting times between successive realizations of the recurrent events. The first approach is most suitable when individuals frequently experience the event of interest and the occurrence does not alter the process itself, while the second approach is more appropriate when the events are relatively infrequent, when, after an event ,individual renewal takes place in some way, or when the focus of the analysis is the prediction of the time to the next event.  For a detailed description of the principles and modelling strategies behind these approaches see 
\cite{CooLaw07}.  In what follows we use both gap and waiting times to indicate the time interval between successive events. 

This paper lies within the waiting times approach and develops a Bayesian semiparametric model for gap times between events.  We assume that the joint distribution of the finite sequence of gap times for each individual is the product of the conditional distributions of each gap time, given the previous ones. 
Moreover, we specify a regression model for each of these conditional distributions to link the realization of each gap time to possibly time-varying covariates and previous waiting times. To account for inter-subject variability,  we introduce individual specific random effects which we model flexibly using a Dirichlet process mixture prior  as random effect distribution. Dirichlet process mixture (DPM) models \citep{Antoniak74, Lo84} are arguably the most common nonparametric Bayesian prior and have proved successful in many applications due to their flexibility 
and ease of  computation.   DPM models are mixtures of a parametric distribution where the mixing measure is the Dirichlet process (DP) introduced by \cite{fer73}.  
It is well known that the DP is
almost surely discrete, and that if $G$ is a DP$(M,G_0)$ with total mass parameter $M$ and baseline distribution $G_0$, then $G$ can be represented as \citep{Sethur94}
\begin{equation*} 
G(\cdot)=\sum_{h\geq 1} w_h \delta_{\theta_h}(\cdot)
\end{equation*}
where $\delta_\theta$ is a point-mass at $\theta$, the weights follow a stick-breaking process, $w_h=V_h\prod_{j<h}(1-V_j)$, with $V_h\iid \textrm{Beta}(1,M)$, and the atoms $\{\theta_h\}_{h\geq 1}$
are such that $\theta_h\iid G_0$.  As the discreteness of $G$ is inappropriate in many applications, it is common to convolve a parametric kernel $k(y\mid \theta) $ with respect to $G$, obtaining a DPM:
$$ H(y)=\int k(y\mid \theta) G(\mbox{d} \theta).$$    
% CITAZIONI : \cite{Lo84}, \cite{Ferguson74},  \cite{Antoniak74}.
\cite{KleIbr98} were the first to adopt a Bayesian nonparametric distribution for the random effects, while in  \cite{MulRos97} we can find one of the earliest examples of the use of  DPM to model  random effects.  \cite{PenDun06}
employ  a Dirichlet process prior to build semiparametric  dynamic frailty models for multiple event time data, allowing also the frailty parameter to change over time.

Due to the discreteness of $G$, 
the DPM prior induces a cluster of the subjects in the sample based on the trajectory of the recurrent events over time, where the number $K$ of clusters is unknown and learned from the data. We investigate different strategies to link gap times at time $t$ with previous gap times. We start by
assuming a standard Markov model where also the order of dependence $p$  is unknown and object of inference.    
We explore two different strategies to specify a prior distribution on $p$: one involves eliciting a prior directly on the space of all possible Markov models for $p\in \{0,1,\ldots, P\}$, while the other approach employs 
spike and slab priors and it is in the spirit of stochastic search variable selection \citep{GeoMc93}.  

In Section \ref{sec:2} we introduce the model, while in Section \ref{sec:3} we explain how to perform inference on the order of dependence in the Markov structure. In Section \ref{sec:simdata} we investigate the performance of the proposed approach in simulations and compare the different strategies to model time dependency and to select the order $p$.   Section \ref{sec:realdata_hosp} and \ref{sec:realdata_UTI}
present two medical applications 
involving recurrent hospitalizations and urinary tract infections, respectively. We conclude the paper in Section \ref{sec:concl}.

\section{Autoregressive random-effects models via Dirichlet process mixtures}
\label{sec:2}

We consider data on $N$ individuals. 
%For a single recurrent event process, starting at time $0$, let $0:=T_{i0}<T_{i1}<\cdots < T_{i n_i}\leq \tau_i$ denote the event times, where $T_{ik}$ is the time of the $k^{th}$ event for individual $i$, $i=1, \ldots,N$.  Let $W_{ij}$, $j=1,2,3,\ldots$ denote the waiting times (gap times) between events of item $i$.  
We assume that $0:=T_{i0}$ corresponds to the start of the event process and that individual $i$ is observed over the time interval $[0,\tau_i]$. If $n_i$ events are observed at times $0<T_{i1}<\cdots < T_{i n_i}\leq \tau_i$, let $W_{ij}=T_{ij}-T_{i j-1}$ for $j=1,\ldots,n_i$ denote the waiting times (gap times) between events of subject $i$ and $W_{i n_i+1}= \tau_i-T_{i n_i}$. Note that if $\tau_i$ corresponds to an event, than 
 $W_{i n_i+1}=0$, while, if it corresponds to end of the observation period, then $\tau_i$ becomes a censoring time. 
Theerefore  $W_{ij}$, $j=1,\ldots, n_i$ are the observed gap times for individual $i$ with a  possible censoring time $W_{i n_i+1}$. 
Let $J$ be the maximum number of observed repeated events, i.e. $J=\max_{i=1,\ldots,N} (n_i)$ and let  $Y_{ij} = \log (W_{ij})$. 
We describe the joint distribution $(Y_{i1}, \dots , Y_{in_i},Y_{in_i+1})$ through the specification of the conditional laws $\mathcal{L}(Y_{ij} | \bm{x}_{ij}, Y_{i1}, \dots , Y_{ij-1})$, where  ${\bm x}_{ij}$ denotes the vector of possibly time-varying covariates for the $i$th individual. In particular, we assume that 
an observation at time $j$, for each subject $i$, $i=1,\ldots,N$ and $j=1,\ldots,n_i$,  is distributed as follows
\begin{equation}
\label{eq:meanY}
%\begin{split}
%Y_{i1}  &= {\bm x}_{i1}^T {\bm\beta}_1+ \sigma \eps_{i1}\\
 Y_{ij} = {\bm x}_{ij}^T {\bm\beta}_j + \alpha_{ij} +\sigma \eps_{ij}, \quad  \eps_{ij}\iid  \Nc(0,1)
%\end{split}
\end{equation}
%and $\eps_{ij}\iid  \Nc(0,1)$ for all $i$ and $j$. 
where ${\bm\beta}_j$ is the vector of regression coefficients at time $j$ common to all individuals. Covariates and regression parameters here have dimension $q$. Moreover, the random parameters $\alpha_{ij}$'s represents time-specific random effects, for which we assume a nonparametric prior with a time-dependent modeling structure as described in subsections \ref{subs:ar1} and \ref{subs:arp}. Given the parameters in the model, the individual recurrent processes are assumed conditionally independent. Note that the number of recurrent events does not need to be the same for all individuals  and that missing data are at least in principle easily accommodated in a Bayesian framework by assuming missingness at random. 

The likelihood for all the sample is then given by:  
$$L=\prod_{i=1}^N \left\{ \left(\prod_{j=1}^{n_i}f(y_{ij}| {\bm z}_{ij}, {\bm\beta}_j, \alpha_{ij},\sigma) \right) S^{\nu_i}(y_{i n_i+1}| {\bm z}_{i n_i+1}, {\bm\beta}_j, \alpha_{ij},\sigma) \right\}$$
where ${\bm z}_{ij}=({\bm x}_{ij},w_{i1},\ldots, w_{i j-1},)$,  $f$ is the density of the gap times (in this case a Gaussian density), $S$ denotes the survival function of the last (censored) gap times and $\nu_i$ is the censoring indicator equal 1 if the last observation is censored.

The vector ${\bm x}_{ij}$ can contain both time-varying and fixed covariates 
and the effect of the covariates can be assumed to be constant over time if appropriate, i.e. ${\bm\beta}_j=  {\bm\beta}$. The vector ${\bm\beta}_j$ does not include the intercept term, because of identifiability issues with $\alpha_{ij}$. 
 Finally, the model can be generalised to include a subject specific or/and time specific observational variance
$\sigma^2$ and/or different distribution for the gap times.

% For preliminaries and notation, see \cite{CooLaw07}, Chapter~4 and Section~2.3. 
%%%%%%%%%%%%%%%%%%%%%%%%%%%%%%%%%%%%%%%%%%%%%%%%%%%%%%%%%%%%%%%%%%%%%%%%%%%%%%%%%%%%%%%%%%%%%%%%%%%%%%%%%%%%%%%%%%%%%%%%%%%%
%%%%%%%%%%%%%%%%%%%%%%%%%%%%%%%%%%%%%%%%%%%%%%%%%%%%%%%%%%%%%%%%%

\subsection{Nonparametric AR(1)-type models}
\label{subs:ar1}
Following a similar  modelling strategy to the one described in 
\cite{DiLucca_etal13}, a straightforward way to introduce dependence among random effects at different times is to allow the distribution of $\alpha_{ij}$ to depend on some summary of the observations up to time $j-1$:
\begin{align}
\label{eq:rand_eff}
\alpha_{ij} \mid m_{i0}, m_{i1} ,\tau &\ind \Nc(m_{i0}+m_{i1} \, f(Y_{i1},\ldots,Y_{i j-1}) , \tau^2), \ \ j=1,\ldots, n_i\\
(m_{i0},m_{i1}) \mid G &\iid  G, \quad G \sim  DP(M,G_0).
\label{eq:DP2}
\end{align}
When $j=1$,  the distribution of the random effect $\alpha_{i1}$ simplifies as the  
autoregressive term in \eqref{eq:rand_eff} disappers and it  reduces to the  Normal distribution with mean $m_{i0}$.
 
%\textcolor{red}{COMMENTS on the model we assume for the random effects. Dependent  DPM etc, single-p DDP, etc.}
We assume conditional independence among subjects, given the parameters, and that  $(m_{i0},m_{i1})$ are independent under the base measure  $G_0$, which becomes the product of a Normal density for $m_{i0}$ and a rescaled Beta for the autoregressive coefficient $m_{i1}$. 
The prior specification  is completed as follows:
%, assuming prior independence among the different blocks of parameters:
\begin{align}
 &{\bm\beta}_j \iid   \Nc_q(0,\beta_0^2 I_q) \nonumber\\%\label{eq:priorbeta}
&\sigma^2 \sim \mbox{Inv-Gamma}(a_{\sigma}, b_{\sigma})\nonumber\\%\label{eq:priorsigma}\\
&\tau^2 \sim \mbox{Inv-Gamma}(a_\tau, b_\tau) \label{eq:priorAR1} \\%\label{eq:priortau}\\ 
&M \sim \Uc(0,M_0) \nonumber\\%\label{eq:priorM}\\
G_0  &= \Nc(0,\sigma_g^2)  \times   \textrm{TBeta}(a_Z,b_Z).\nonumber %\label{eq:G0}
\end{align}
By TBeta$(a_Z,b_Z)$ we mean the translated Beta distribution defined on the interval $(-1,1)$ with density proportional to ${\displaystyle (y+1)^{a_Z-1}}(1-y)^{b_Z-1}$ $\Uno_{(-1,1)}(y)$.  
The prior distribution on $\sigma$ and $\tau$ can be replaced by a uniform distribution with a large support as this strategy allows for better computations when using Bayesian softwares such as JAGS. 
We constrain  the support of the marginal  distribution of $m_{i1}$, as in the Gaussian AR(1) model,  to be  $(-1,1)$ since conditionally on $\mathbf{\theta}_i=(m_{i0},m_{i1},\sigma^2,\tau^2)$, the distribution of  $\alpha_{ij}$ is Gaussian with parameters 
\begin{eqnarray}
\E (\alpha_{ij}\mid \mathbf{\theta}_i) &=& m_{i0}\left(1+m_{i1} +\cdots + m_{i1}^{j-2}\right)
\nonumber
\\ \nonumber 
\Var  ( \alpha_{i2}\mid \mathbf{\theta}_i) &=&  \tau^2+ m_{i1}^2 \sigma^2\\
\Var ( \alpha_{ij}\mid \mathbf{\theta}_i) &=&  \tau^2\left( 1+m_{i1}^2 + \cdots + (m_{i1}^2)^{j-2}\right)
\nonumber \\
&\quad+&\sigma^2      m_{i1}^2  \left( 1+m_{i1}^2 + \cdots + (m_{i1}^2)^{j-3}\right),\,  j\geq 3.
\label{eq:infvar}
\end{eqnarray}
The above equations are easily obtained marginalising over the distribution of $\mathbf{Y}_i$ and ignoring the covariate term. From \eqref{eq:infvar} it is evident that if $|m_{i1}|\geq 1$, the variance of  $\alpha_{ij}$ tends to infinity as $j$ increases, leading to a non-stationary process. Therefore, constraining the support to be $(-1,1)$ leads to more stable computations. 
%So the condition on the support of $m_{i1}$ guarantees (????) the weakly stationarity of the time series.

The choice of $f$ is obviously crucial  and  depends on the context and the goals of the inference problem. Common alternatives in the literature are:
\begin{itemize}
\item $f(Y_{i1},\ldots, Y_{i j-1})= Y_{i j-1}$, i.e. the random effect at time $j$ has a Dirichlet process mixture prior, where the location points are modeled as a AR(1) model - that is the observation at time $j-1$ influences the behaviour of the random effect at time $j$;
\item $f(Y_{i1},\ldots, Y_{i j-1})= (Y_{i 1}+\cdots+ Y_{i j-1})/(j-1)$, i.e.  conditional expected value of each $\alpha_{ij}$ depends on the sample mean of the observations up to time  $j-1$; 
%or the geometric mean of the sample $ (Y_{i 1}\times\cdots\times Y_{i j})^{1/k}$. 
\item $f(Y_{i1},\ldots,Y_{i j-1})= \left(Y_{i1}\times \cdots\times Y_{i j-1} \right)^{1/(j-1)}$; this is equivalent to the geometric mean of the observations up to time  $j-1$.
 %$m_{i0}+m_{i1}  \log \left( W_{i 1}+\cdots+ W_{i j-1}\right)/(j-1)$ (sample mean of the previous gap times) 
\end{itemize}
%Obviously if $T_{iJ^o} $ is the last observed gap time for indivdual $i$, we have $ T_{iJ^\star} >  T_{iJ^\star -1} > \cdots > T_{iJ^o} $.

Note that, when $f(Y_{i1},\ldots, Y_{i j-1}) =Y_{i j-1}$, then \eqref{eq:rand_eff}-\eqref{eq:DP2} imply that the random effects distribution at time $j$ is a DPM of AR(1) processes, with dependence only on the gap time at time $j-1$. 
%COMPARISON to SIMILAR MODELS:  ma QUALI? \cite{PenDun06}? Ma è un po' diverso dal nostro.  
%Citare \cite{KleIbr98} (questo è il primo lavoro, in cui i random effects sono iid da DP),  \cite{MulRos97} and \cite{MukGel97} tra i primi paper con random effects iid da un DPM, 

%%%%%%%%%%%%%%%%%%%%%%%%%%%%%%%%%%%%%%%%%%%%%%%%%%%%%%%%%%%%%%%%%%%%%%%%%%%%%%
%%%%%%%%%%%%%%%%%%%%%%%%%%%%%%%%%%%%%%%%%%%%%%%%%%%%%%%%%%%%%
\subsection{Nonparametric AR(p) Models}
\label{subs:arp}
The model in Subsection~\ref{subs:ar1} can be extended to include higher order dependence, by modifying  \eqref{eq:rand_eff} -\eqref{eq:DP2} as follows:
\begin{align}
\label{eq:randarp}
\alpha_{ij} \mid m_{i0}, m_{i1} ,\ldots, m_{ip}, \tau &\ind \Nc(m_{i0}+\sum_{l=1}^p m_{il} \, Y_{ij-l}, \tau^2), \ \ j=p+1,\ldots, n_i\\
(m_{i0},m_{i1},\ldots,m_{ip} ) \mid G \iid  G, &\quad G \sim  DP(M,G_0)
\label{eq:mult_m}\\
G_0  = \Nc(0,\sigma_g^2)  \times  & \underbrace{\textrm{TBeta}(a_Z,b_Z) \times \cdots \times \textrm{TBeta}(a_Z,b_Z)}_{p \mbox{ times}} 
\label{eq:p-dimG0}
\end{align}
%\textcolor{red}{BE CAREFUL: this is not the largest support that guarantees stationarity or whatever for AR(p) models! Should we care about that?}

The distribution of $\alpha_{ij}$ for $j \leq p $, depends only on the available past observations as in any AR$(p)$ model.
%The prior for the remaining parameters  is similar   as given in the previous subsection.
%%%%%%%%%%%%%%%%%%%%%%%%%%%%%%%%%%%%%%%%%%%%%%%%%%%%%%%%%%%%%%%%%%%

\section{Testing for the Order of Dependence}
\label{sec:3}
In \eqref{eq:randarp} we assume that the order of dependence on past observations is a fixed integer $p$. 
However, this parameter is often unknown in applications, and it needs to be estimated. A wealth of research focuses on Bayesian model selection
\citep[see][for example]{GeoMc97, ClyGeo04}. 
Here we concentrate on two approaches.  The first one modifies the base measure of the DP by including a spike and slab distribution on the autoregressive coefficient, leading to Spiked Dirichlet process prior introduced by  \cite{kim2009spiked}.
The second one involves the direct specification of a prior on $p$, and then, conditional on $p$, we specify the prior distribution for the remaining parameters; in this case the dimension of the parameter vector $(m_{i0}, m_{i1} ,\ldots, m_{ip})$ changes 
according to $p$ and consequently the dimension of the space where the Dirichlet process measure is defined.

\subsection{Spike and slab Variable Selection}
\cite{kim2009spiked} introduce Spiked Dirichlet process prior in the context of regression.  A key feature of their method is to employ a spike and slab distribution, i.e. a mixture of a point mass at 0 and a continuous distribution 
as centering distribution of the DP. This implies that, in a regression context, some coefficients have a positive probability of being equal to 0 and therefore not influential on the response of interest. Their technique is 
easily accommodated in our context by simply modifying $G_0$ in \eqref{eq:p-dimG0} as
\begin{eqnarray}
G_0 &=& \Nc(0,\sigma_g^2)  \times   \underbrace{\pi_1(a_Z,b_Z) \times \cdots \times \pi_p(a_Z,b_Z)}_{p \mbox{ times}} \nonumber \\
\pi_l(a_Z,b_Z) &=& (1-\eta_l)\delta_0+\eta_l \textrm{TBeta}(a_Z,b_Z), \ l=1,\ldots,p  \label{eq:spikeslabG0}\\
 \eta_l &\ind& \textrm{Bernoulli}(c_l)\nonumber \\
 c_l &\iid& \Uc(0,1) \nonumber
%\label{eq:spikeslabmix}
\end{eqnarray}
where the introduction of hyperpriors on the weights of the mixture induces sparsity. 
%and then introducing hyperpriors on the weights of the mixture
%\begin{eqnarray}
% \eta_l &\ind& \textrm{Bernoulli}(c_l) \\
% c_l &\iid& \Uc(0,1)
%\label{eq:spikeslabbern}. 
%\end{eqnarray}

\subsection{Prior on the Order of Dependence}
\label{subs:order}
Following \cite{QuiMul_orddep}, we specify a prior directly on the order $p$ of the autoregressive process and then, conditioning on $p$, we set a Dirichlet Process prior of appropriate dimension for the parameters of the AR(p), i.e. the vector $(m_{i0},m_{i1},\ldots,m_{ip} )$. Let $P$ be the maximum possible order. Then we can specify the following hierarchy: 
\begin{eqnarray}
\alpha_{ij} \mid p,  m_{i0}, m_{i1} ,\ldots, m_{ip}, \tau &\ind&  \Nc(m_{i0}+\sum_{l=1}^p m_{il} \, Y_{ij-l}, \tau^2), \quad j=p+1,\ldots, n_i  \nonumber\\
(m_{i0},m_{i1},\ldots,m_{ip} ) \mid p, \widetilde G_p &\iid&  \widetilde G_p \nonumber\\
 \widetilde G_p &\sim&  DP(M,G_{0p}) \label{eq:randarp2}\\
%\label{eq:multiDP}  \\
G_{0p}  &= & \Nc(0,\sigma_g^2)  \times   \underbrace{\textrm{TBeta}(a_Z,b_Z) \times \cdots \times \textrm{TBeta}(a_Z,b_Z)}_{p \mbox{ times}} \nonumber\\
%\label{eq:p-dimG0l}\\
p &\sim &\mbox{ Discrete Uniform on  } \{0,1,\ldots,P \} \nonumber
%\label{eq:p-dimG0l_p}
\end{eqnarray}
When $p=0$, the process simplifies as the  
autoregressive term in \eqref{eq:randarp2} disappers and the base measure of the DP reduces to the  Normal distribution for the intercept term. 
%Moreover,  the $p+1$ Dirichlet processes $\widetilde G_0$, $\widetilde G_1$, $\ldots$, $\widetilde G_p$ are a priori independent, conditionally in $l$. 

%%%%%%%%%%%%%%%%%%%%%%%%%%%%%%%%%%%%%%%%%%%%%%%%%%%%%%%%%%%%%%%%
%\newpage
\section{Simulated data}
\label{sec:simdata}

In order to check the performance of the class of models proposed in the previous sections, two different simulated scenarios have been created. 
%Here the results of tests for the order of dependence are illustrated.
Posterior inference for these examples, as well as for the real data applications in Section~\ref{sec:realdata_hosp} and \ref{sec:realdata_UTI}, can be performed through a standard Gibbs sampler algorithm, which we implement in JAGS \citep{Plummer03jags},  using a truncation-based algorithm 
for stick-breaking priors \citep{IshZar2002exact}. 
For all simulations, we run the program for $251,000$ iterations, discarding the first $1,000$ iterations as burn-in and thinning every $50$ iterations to reduce the autocorrelation of the Markov chain. 
The final sample size is then $5,000$. Unless otherwise stated, we check through standard diagnostics criteria such as those available in the R package CODA \citep{CODA} that convergence of the chain is satisfactory for most of the parameters.

\subsection{Simulation scenario 1: Spike and slab Variable Selection}
We consider a simulated dataset of $N=300$ subjects, with $n_i=10$ for all $i$. One third of the observations are generated from
\begin{eqnarray*}
Y_{ij} \sim \Nc(0, (1.2)^2), \ j=1,\ldots, 10
\end{eqnarray*}
while another third is  generated from 
\begin{eqnarray*}
 Y_{i1} \sim \Nc(0, (1.5)^2), &\quad& Y_{i2}| Y_{i1}  \sim \Nc( Y_{i1}, (1.5)^2)\\
 Y_{ij}| Y_{ij-1},Y_{ij-2}  &\sim& \Nc( Y_{ij-1} + 0.7 \times Y_{ij-2}, (1.5)^2), \ j=3,\ldots, 10
\end{eqnarray*}
and the last 100 observations are generated  from
\begin{eqnarray*}
Y_{i1} \sim \Nc(0, (0.9)^2), &\quad& Y_{i2}| Y_{i1}  \sim \Nc( Y_{i1}, (0.9)^2) \\
Y_{i3}| Y_{i2},Y_{i1}  &\sim& \Nc( Y_{i2} + 0.7 \times Y_{i1}, (0.9)^2) \\
Y_{ij}| Y_{ij-1},Y_{ij-2},Y_{ij-3}  &\sim& \Nc( Y_{ij-1} + 0.7 \times Y_{ij-2}+0.4 \times Y_{ij-3}, (0.9)^2), \ j=4,\ldots, 10.
\end{eqnarray*}
In simulating the data, we assume independence across subjects. In this example, for ease of explanation, we do not include covariates.  
%The covariates are not present in the generating model, i.e. ${\bm x}_{ij}=0$ for all $i$ and $j$. 
%In Figure~\ref{fig:datasim1} the scatterplots of the observations are displayed, colored according to their group. Let us remark that $J=10$. However, we here decided to illustrate only the first five times for simplicity.
%As one can see, in the first group of patients, represented in red, there is not correlation between $Y_{ij}$ at different  time $j$. Instead, in the second and in the third group a positive correlation is evident.
%\begin{figure}[!ht]
%\centering
%\includegraphics[width=0.32\columnwidth]{Images/sp1.png}
%\includegraphics[width=0.32\columnwidth]{Images/sp2.png}
%\includegraphics[width=0.32\columnwidth]{Images/sp3.png}
%\includegraphics[width=0.45\columnwidth]{Images/readmission_m3_SSVS.png}
%\caption{Posterior predictive  distribution of $(m_{i0},m_{i1},m_{i2},m_{i3})$.}
%\label{fig:pred_m_SSVS}
%\end{figure}
%\caption{Scatterplot of the first simulated dataset; the colors indicate the three different groups.}
%\label{fig:datasim1}
%\end{figure}

We fit the model \eqref{eq:meanY}, \eqref{eq:randarp}-\eqref{eq:mult_m}, where $G_0$ is given by the product of spike and slab distributions as defined in 
\eqref{eq:spikeslabG0}. %-\eqref{eq:spikeslabbern}. 
In fitting the model we set $p=3$ and 
\begin{eqnarray*}
\sigma_g^2 &=& 10, \quad a_Z = 3, \quad b_Z = 3\\
\sigma &\sim& \Uc(0,10)\\
\tau &\sim& \Uc(0,10)\\ 
M_0 &=& 10.
\end{eqnarray*}
Hyperparameters are chosen in order to specify vague marginal prior distributions. 
%related to the base measure $G_0$ of the DP prior are chosen as non-informative as possible. This choice of the priors is made in order to specify vague prior distributions. 

\begin{figure}[h!]
\centering 
\subfigure[]%
{\includegraphics[width=0.40\textwidth,height=0.355\textwidth]{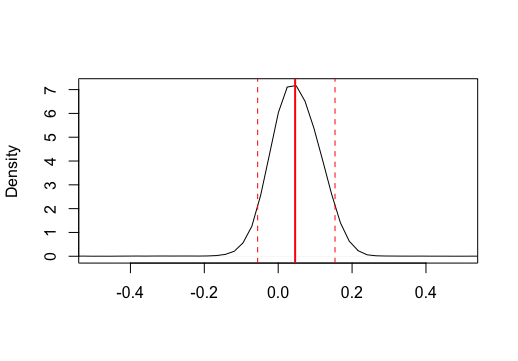}} 
\subfigure[]%
{\includegraphics[width=0.40\textwidth,height=0.35\textwidth]{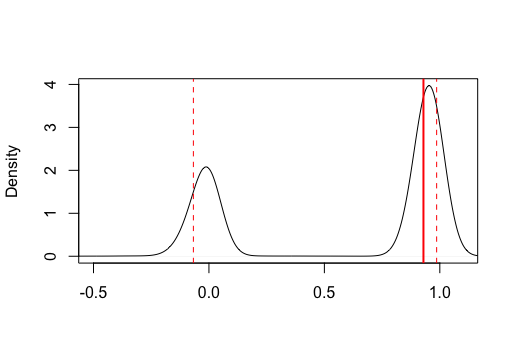}} 
\subfigure[]%
{\includegraphics[width=0.40\textwidth,height=0.35\textwidth]{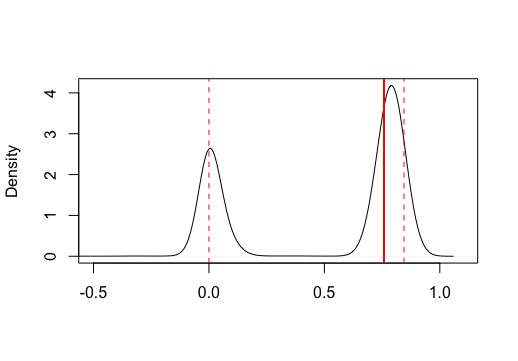}}  
\subfigure[]%
{\includegraphics[width=0.40\textwidth,height=0.35\textwidth]{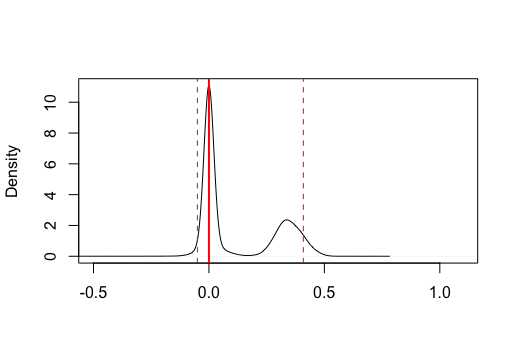}} 
\caption{Simulation scenario 1: predictive marginal distributions of  $m_{i0}$(a), $m_{i1}$(b), $m_{i2}$(c) and $m_{i3}$(d) .  Dashed vertical lines denote 0.05 and 0.95 posterior quantiles, while the bold vertical line indicates the posterior median.}
\label{fig:simul1_post_m}
\end{figure}

Figure~\ref{fig:simul1_post_m} shows the predictive distributions of $m_{i0}$, $m_{i1}$, $m_{i2}$ and $m_{i3}$.
By visual inspection, it is clear that the results of  the predictive distributions of $m_{ij}$ agree with the true values used to create the dataset. In fact, the predictive distribution of $m_{i0}$ is concentrated around $0$, while the predictive distributions of $m_{i1}$, $m_{i2}$ and of $m_{i3}$ are bimodal with mode around  $\{0, 1\}$, $\{0, 0.7\}$ and $\{0, 0.4\}$, respectively.
%Moreover, the posterior marginal distributions of  $\eta_{1}$ and $\eta_{2}$ concentrate most mass on $1$, whereas the posterior marginal distribution of $\eta_{3}$ is more concentrated on 0 than on 1.

\begin{figure}[!ht]
\centering
\includegraphics[width=0.4\columnwidth]{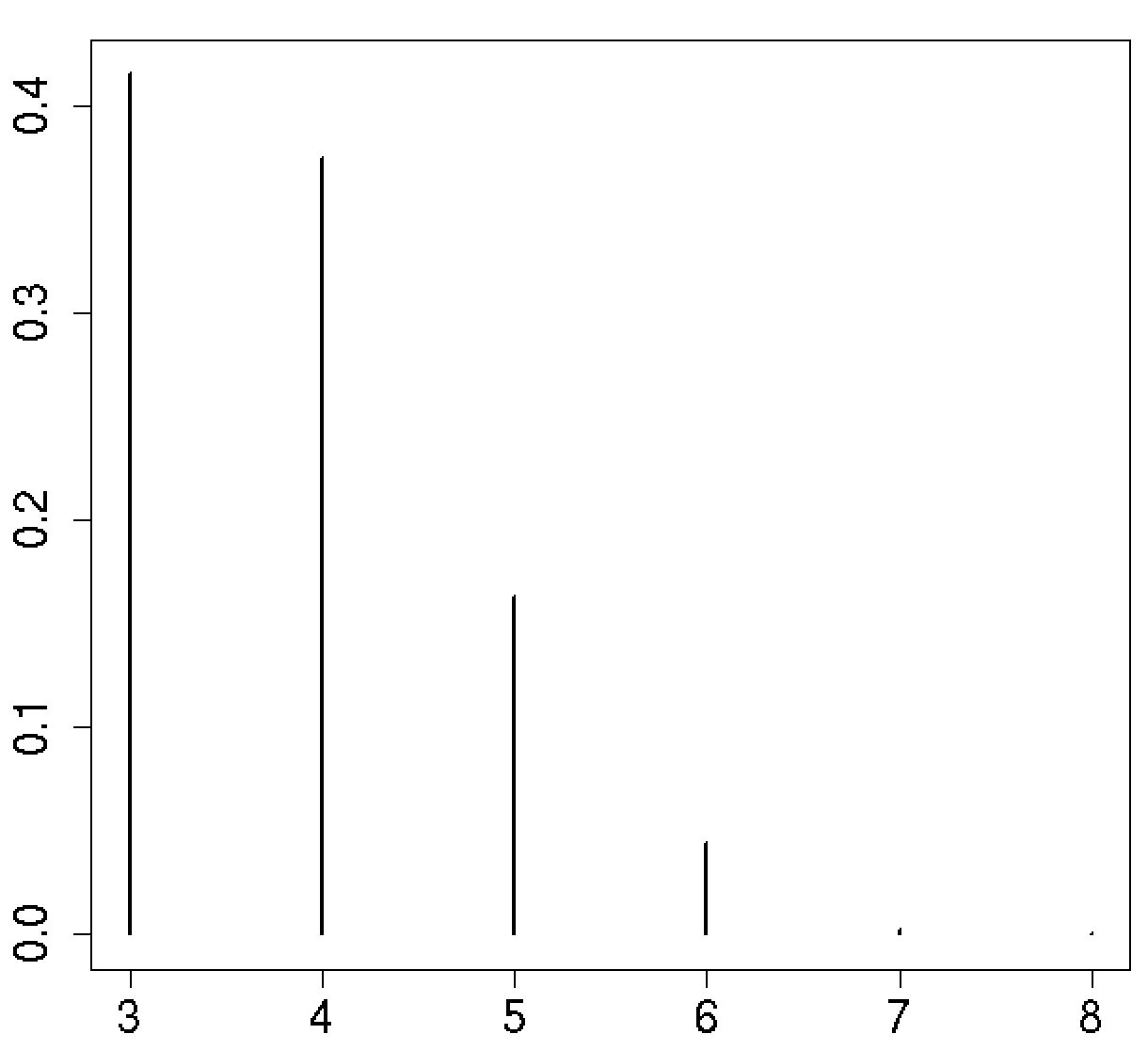}
\caption{Simulation scenario 1: posterior distribution of $K$.}
\label{fig:simul1_post_K}
\end{figure}

%\begin{figure}[!ht]
%\centering
%\includegraphics[width=0.8\columnwidth,height=0.3\textwidth]{Images/sim1_eta-eps-converted-to}
%%\includegraphics[width=\textwidth,height=0.3\textwidth]{Images/sim1_eta.eps}
%%\includegraphics[width=0.99\columnwidth]{Images/sim1_eta}
%\caption{First simulated dataset: posterior marginal distributions of  $\eta_{1}$, $\eta_{2}$ and $\eta_{3}$.}
%\label{fig:simul1_post_eta}
%\end{figure}
%As one can see from Figure~\ref{fig:simul1_post_eta}, 
The marginal posterior distributions of $\eta_1$ and $\eta_2$, not reported here, concentrate most mass on $1$, with posterior probability of being equal to 1 of approximately 0.8 and 0.75, respectively. The marginal  posterior distribution of $\eta_3$ shows more uncertainty, with posterior probability of being equal to 1 close to 0.44.
These results capture the data generating process as 
200 observations have a temporal dependency of the second order  and 100 observations have a dependency of the third order. Moreover,
Figure~\ref{fig:simul1_post_K} displays the predictive distribution of $K$, the number of distinct components in the mixture \eqref{eq:randarp}-\eqref{eq:mult_m}.
The configurations involving 3 or 4 clusters are clearly those with the highest posterior probability: posterior inference on $K$ is in agreement with the 3 components  used to generate the data.

\subsection{Simulation Scenario 2: Prior on the Order of Dependence}
In this section we simulate a dataset of $N=200$, with $n_i=10$ for all $i$. Half observations are generated independently from
\begin{eqnarray*}
 Y_{i1} \sim \Nc(0, 1.5^2), &\quad& Y_{i2}| Y_{i1}  \sim \Nc( 0.9 \times Y_{i1}, 0.9^2)\\
 Y_{ij}| Y_{ij-1},Y_{ij-2}  &\sim& \Nc( 0.9 \times Y_{ij-1} + 0.7 \times Y_{ij-2}, 0.9^2), \ j=3,\ldots, 10
\end{eqnarray*}
while the other half is independently generated from
\begin{eqnarray*}
 Y_{i1} \sim \Nc(0, 1.5^2), &\quad& Y_{i2}| Y_{i1}  \sim \Nc( -0.9 \times Y_{i1}, 1.5^2) \\
 Y_{ij}| Y_{ij-1},Y_{ij-2}  &\sim& \Nc( -0.9 \times Y_{ij-1} - 0.7 \times Y_{ij-2}, 1.5^2), \ j=3,\ldots, 10
\end{eqnarray*}
As in the previous example,  covariates are not present in the generating model.
% Figure~\ref{fig:datasim2} shows the scatterplots of the observations $\{y_{ij}, i \}$, for $j=1,\ldots,5$,   
% colored according to their group. %As in the previous simulation, we decided to illustrate only the first five times.
% \begin{figure}[!ht]
% \centering
% \includegraphics[width=0.45\columnwidth]{Images/sp1_2.png}
% \includegraphics[width=0.45\columnwidth]{Images/sp2_2.png}
% \caption{Scatterplot of the second simulated dataset; the colors indicate the different groups.}
% \label{fig:datasim2}
% \end{figure}
% As one can see, in the first group of patients, represented in red, there is a positive correlation between $Y_{ij}$ at different  time $j$, whereas in the second group there is a negative correlation.
% Moreover, is evident that in the second group of observations there is more dispersion because we generated the observations $Y_{ij}$ with a bigger variance. 

We fit model  \eqref{eq:meanY}, \eqref{eq:randarp2} %-\eqref{eq:p-dimG0l_p} 
to this dataset, with maximum  order of dependence  $P=3$ and prior hyperparameters (corresponding to a vague prior) set as follows:
\begin{eqnarray*}
\sigma_g^2 &=& 10, \quad a_Z = 3, \quad b_Z = 3\\
\sigma &\sim& \Uc(0,10)\\
\tau &\sim& \Uc(0,10)\\ 
M &\sim& \Uc(0, 5). 
\end{eqnarray*}

\begin{figure}[h!]
\centering 
\subfigure[]%
{\includegraphics[width=0.40\textwidth,height=0.355\textwidth]{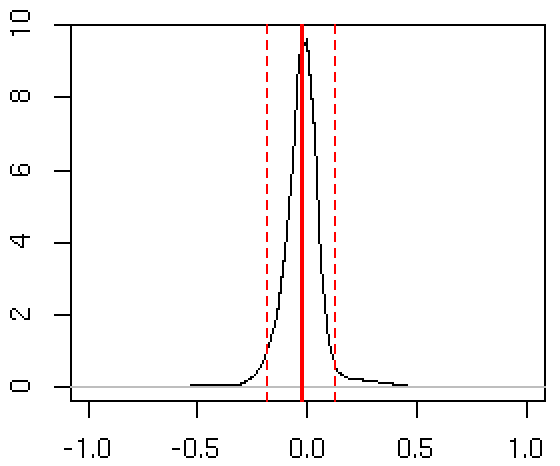}} 
\subfigure[]%
{\includegraphics[width=0.40\textwidth,height=0.35\textwidth]{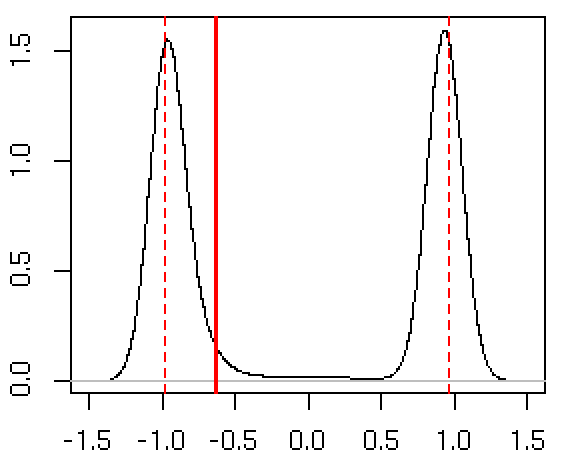}} 
\subfigure[]%
{\includegraphics[width=0.40\textwidth,height=0.35\textwidth]{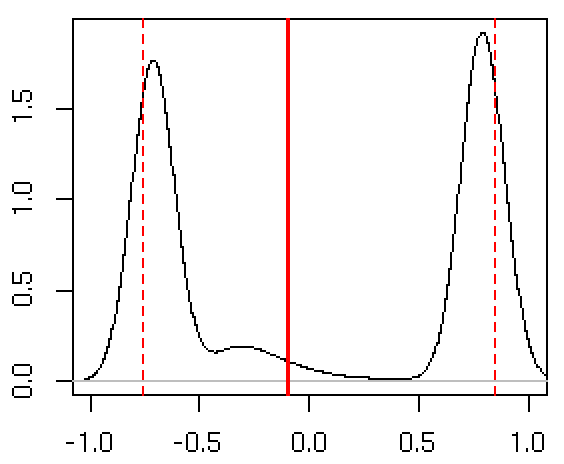}}  
\subfigure[]%
{\includegraphics[width=0.40\textwidth,height=0.35\textwidth]{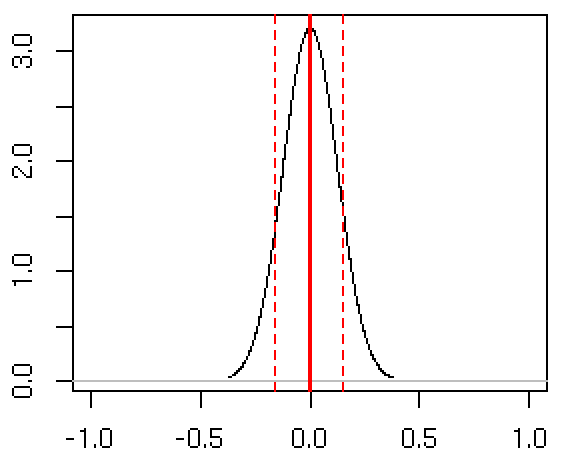}} 
\caption{Simulation scenario 2: predictive marginal distributions of  $m_{i0}$(a), $m_{i1}$(b), $m_{i2}$(c) and $m_{i3}$(d), conditioning on $p=2$.  Dashed vertical lines denote 0.05 and 0.95 posterior quantiles, while the bold vertical line is the posterior median.}
\label{fig:simul2_post_m}
\end{figure}

The mode of the marginal posterior distribution of $p$ is 2, with corresponding posterior probability almost 1. 
Conditional on $p=2$, 
Figure~\ref{fig:simul2_post_m} reports 
%~\ref{fig:simul2_post_p} and 
the predictive distributions of $m_{i0}$, $m_{i1}$, $m_{i2}$ and $m_{i3}$. Once again, the result 
of inference are in agreement with the true parameters used to generate the data, which are realizations of a  second order Markov process.
From Figure~\ref{fig:simul2_post_m} it is evident that the 95\% posterior credible intervals (CIs) for $m_{ij}$, $j=0,1,2,3$,  cover the true values. More in details, 
the predictive distributions of $m_{i0}$ and of $m_{i3}$ are concentrated around $0$, while the predictive distributions of $m_{i1}$ and of $m_{i2}$ are bimodal with  mode  around $\{-0.9, 0.9\}$ and $\{-0.7, 0.7\}$, respectively.
Finally, conditioning on $p=2$, the posterior mode for the number $K$ of clusters is 2, with associated posterior probability equal to 0.5.

\section{Hospitalization dataset}
\label{sec:realdata_hosp} %Sect2
%We apply the model described so far to the context of recurrent events. 
%For the ease of notation, we suppress 
%By renewal processes we mean those ones in which the gap times of each patient $W_{ij}$, $j=1,\ldots,n_i$ are independent and conditionally distributed (iid), conditionally to covariates and parameters. However, this assumption is untenable in most situations. 
%Here we model the more general situation where the joint distribution of each vector $(W_{i1},\ldots,W_{i n_i})$ can be specified through the sequence of conditional distributions 
%\begin{equation}
%\Lc(W_{ij}|x_{ij}, W_{i1},\ldots,W_{i j-1}), \quad j=2,3,\ldots, 
%\label{eq:cond}
%\end{equation}
%in addition to $ \Lc(W_{i1}|x_{i1})$.  
%\textcolor{red}{Pros and cons for conditional models for recurrent events}:  in particular see \cite{CooLaw07}, section 4.2, \cite{ChaWan99}  and \cite{Prentice_etal81}.
%In orther to flexibly model distributions \eqref{eq:cond}, we will assume Bayesian nonparametric mixture models for equations \eqref{eq:cond}. 
We fit model  \eqref{eq:meanY}-\eqref{eq:DP2}   to the  \textit{readmission} dataset in the R package \textit{frailtypack}  for all the possible choices of $f$ described in Section 2.1. The dataset contains rehospitalization times (in days) after surgery in patients diagnosed with colorectal cancer.  
Data are available on $N = 403$ patients, for a total number of $861$ recurrent events. In addition to gap times between successive rehospitalizations, the dataset contains information for each patient on the following covariates:
\begin{itemize}
\item  \textit{chemo}: variable indicating if the patient received chemotherapy.
\item \textit{sex}: gender of the patient.
\item \textit{dukes}: ordinal variable indicating the classification of the colorectal cancer. 
%The cancer is more and more severe as this variable increases: 
The baseline A-B denotes the invasion of the tumor through the bowel wall penetrating the muscle layer but not involving lymph nodes; the value C indicates the involvement of lymph nodes; the value D implies the presence of widespread metastases. Category D corresponds to the most severe type of cancer. 
\item \textit{charlson}: Charlson comorbidity index. 
It measures  ten-year mortality for a patient who may have a range of comorbidity conditions, and 
ranges within 3 classes, i.e. $\{0, 1-2, 3\}$. This is the only time-varying covariate. 
\end{itemize}
The recurrent events in this study are readmission times  (colorectal cancer patients may have several readmissions after first discharge). 
The origin of the time axis is the date of the surgical procedure for each patient and the recurrent events are next rehospitalizations related to colorectal cancer. In the analysis, we consider only patients with 6 or less events, leaving a dataset of $N = 197$ patients and a total number of 495 recurrent events.  Table~\ref{tab:ni_real} reports the number of patients with exactly $j$ gap times, for $j = 1, \dots, 6$. Moreover,  119 observations out of 197 are right-censored with respect to their last gap time.  Since the proportion of censored data is considerably high, we need to take censoring into account.
\begin{table}[!ht]
\centering
\begin{tabular}{|c|c|c|c|c|c|c||c|}
\hline 
$j$ & $1$ & $2$ & $3$ & $4$ & $5$ & $6$ & TOT\\ 
\hline 
$n_j$ & $30$ & $96$ & $36$ & $18$ & $9$ & $8$ & $197$\\ 
\hline 
\end{tabular}
\caption{Number of patients for $j$ gap times, $j = 1, \dots, J$.} 
\label{tab:ni_real}
\end{table}

%In particular, we applied model \eqref{eq:meanY}-%\eqref{eq:rand_eff}\eqref{eq:DP2}, where, in this case 
%\begin{eqnarray*}
%{\bm x}_{ij}^T {\bm\beta}_j = x_{i1}^T {\beta_{1}}+x_{i2}^T {\beta_{2}}+x_{i3}^T {\beta_{3}}+x_{i4}^T {\beta_{4}}+x_{ij5}^T {\beta_{j5}}+x_{ij6}^T {\beta_{j6}};
%\end{eqnarray*}
%the first and second covariates denotes if the patient received chemotherapy ($x_{i1}=1$) and the gender (=1 if woman), while  $(x_{i3}, x_{i4})$ are dummy variables for the ordinal covariate \textit{dukes} (baseline: A-B) and  $(x_{i5}, x_{i6)}$ indicate the  \textit{charlson} covariate (baseline: 0); these last two covariates are time-varying. 

Prior hyperparameters in \eqref{eq:priorAR1} %-\eqref{eq:G0} 
are set as follows:
\begin{eqnarray*}
\beta_0^2 &=& 1,000\\
\sigma &\sim&  \Uc(0,10)\\
\tau &\sim&  \Uc(0,10)\\
%M &=& 1\\
\sigma_g^2 &=& 10, \quad a_Z = 3, \quad b_Z = 3\\ 
M &=& 1 \ . 
\end{eqnarray*}
%This choice is made  in order to specify vague prior distributions. 
%For computational difficulties, we fixed the total mass parameter $M=1$, which corresponds to {\color{red}$\E(K_n)=$ ?}. 

\subsection{Testing for the Order of Dependence}
\label{subs:read_order}
\begin{figure}[h!]
\centering 
%{\includegraphics[width=0.45\textwidth,height=0.4\textwidth]{Images/read_p_order.png}} 
\includegraphics[width=0.45\textwidth,height=0.4\textwidth]{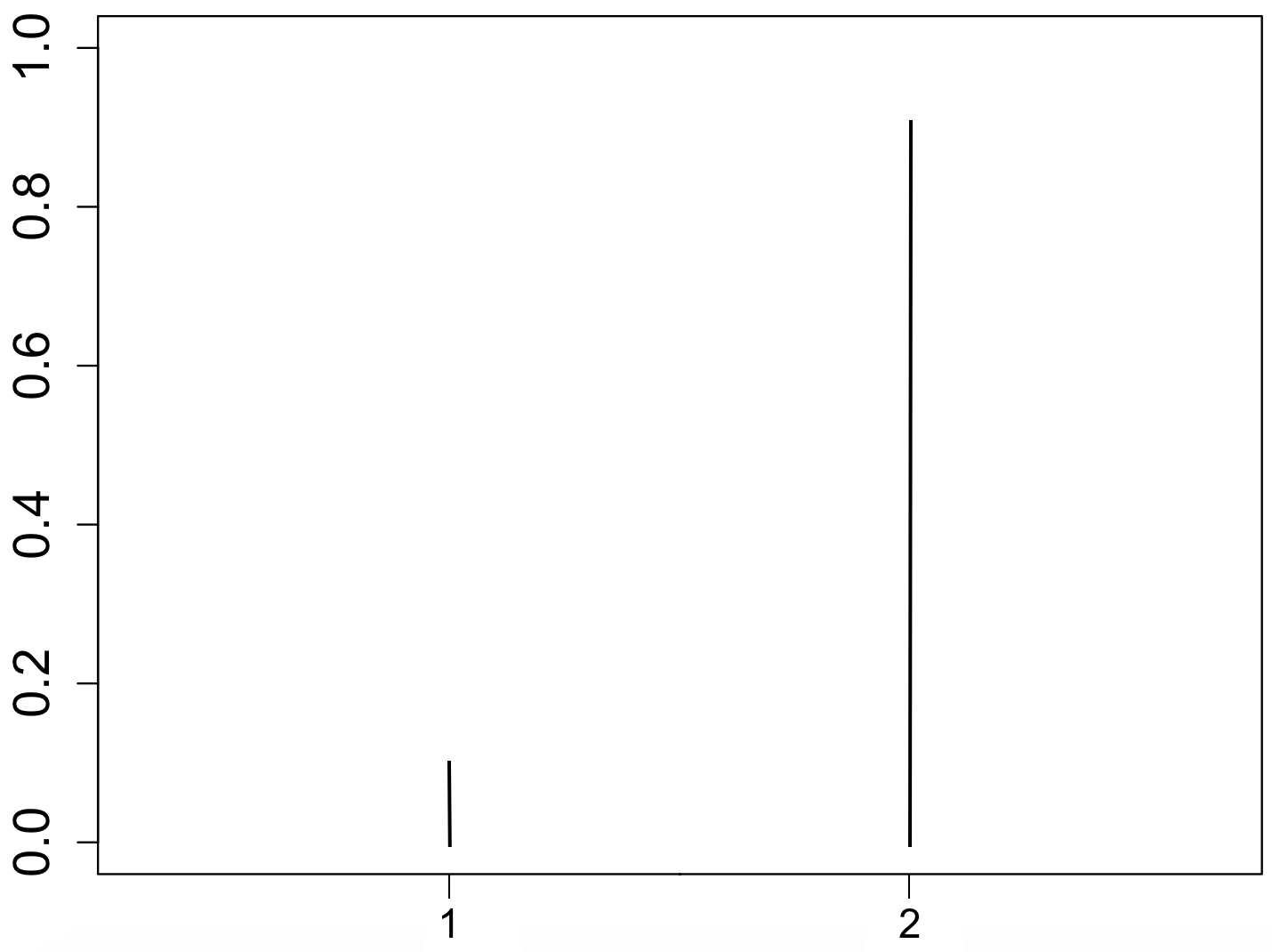}
\caption{ \textit{Readmission} dataset: posterior distribution of $p$.}
\label{fig:read_order_p}
\end{figure}

When testing the  order of dependence, we first fit model \eqref{eq:meanY} and \eqref{eq:spikeslabG0} %-\eqref{eq:spikeslabbern} 
with $p=3$ ($G_0$ being a spike and slab distribution) and then model \eqref{eq:meanY} and prior  \eqref{eq:randarp2} %-\eqref{eq:p-dimG0l_p} 
with $P=3$. 
\begin{figure}[h!]
\centering 
\subfigure[]%
{\includegraphics[width=0.40\textwidth,height=0.35\textwidth]{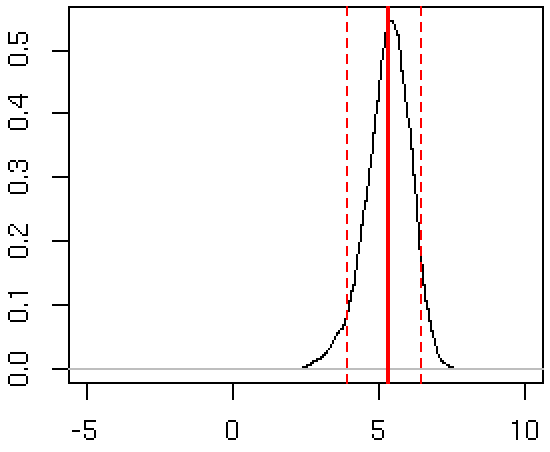}} 
\subfigure[]%
{\includegraphics[width=0.40\textwidth,height=0.35\textwidth]{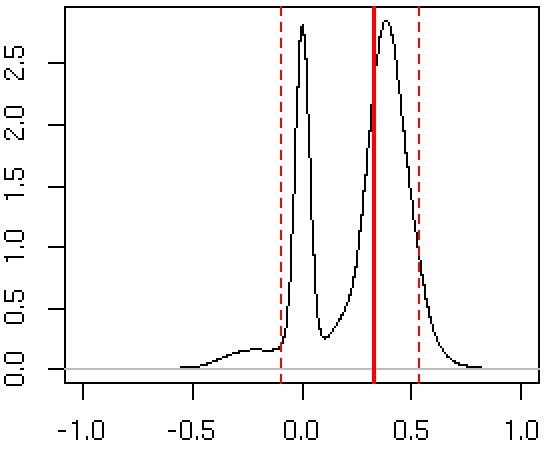}} 
\subfigure[]%
{\includegraphics[width=0.40\textwidth,height=0.35\textwidth]{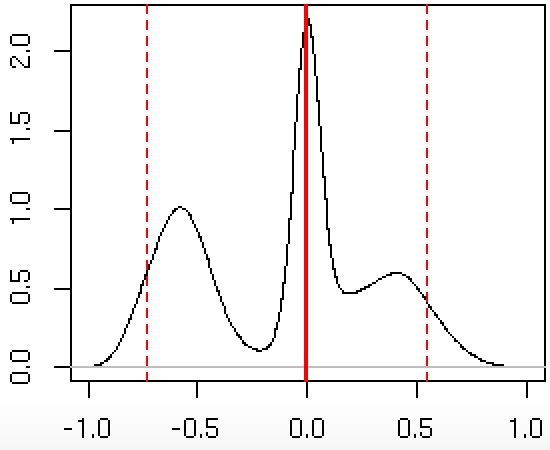}}  
\subfigure[]%
{\includegraphics[width=0.40\textwidth,height=0.35\textwidth]{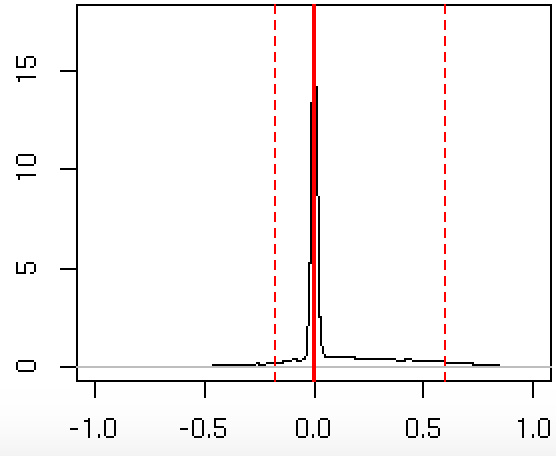}} 
\caption{\textit{Readmission} dataset: predictive marginal distributions of  $m_{i0}$(a), $m_{i1}$(b), $m_{i2}$(c) and $m_{i3}$(d).  Dashed vertical lines denote 0.05 and 0.95 posterior quantiles, while the bold vertical line is the posterior median.}
\label{fig:read_SSVS_m}
\end{figure}
Figure~\ref{fig:read_SSVS_m} reports the posterior predictive marginal distributions of $m_{i,l}$, for $l=0,1,2,3$, obtained with spike and slab variable selection. Since the marginal posterior distributions of $m_{i0}, m_{i1}, m_{i2}$ are not concentrated around 0, unlike that of $m_{i3}$, we can conclude that the process best describing the \textit{readmission} dataset has a dependency of the second order.
This result is confirmed also using the approach described in  Section~\ref{subs:order}. Indeed,
the posterior distribution of $p$, displayed in Figure~\ref{fig:read_order_p}, places most of its mass on $2$.

\subsection{Posterior analysis}
\begin{figure}[h!]
\centering 
\subfigure[]%
{\includegraphics[width=0.325\textwidth,height=0.35\textwidth]{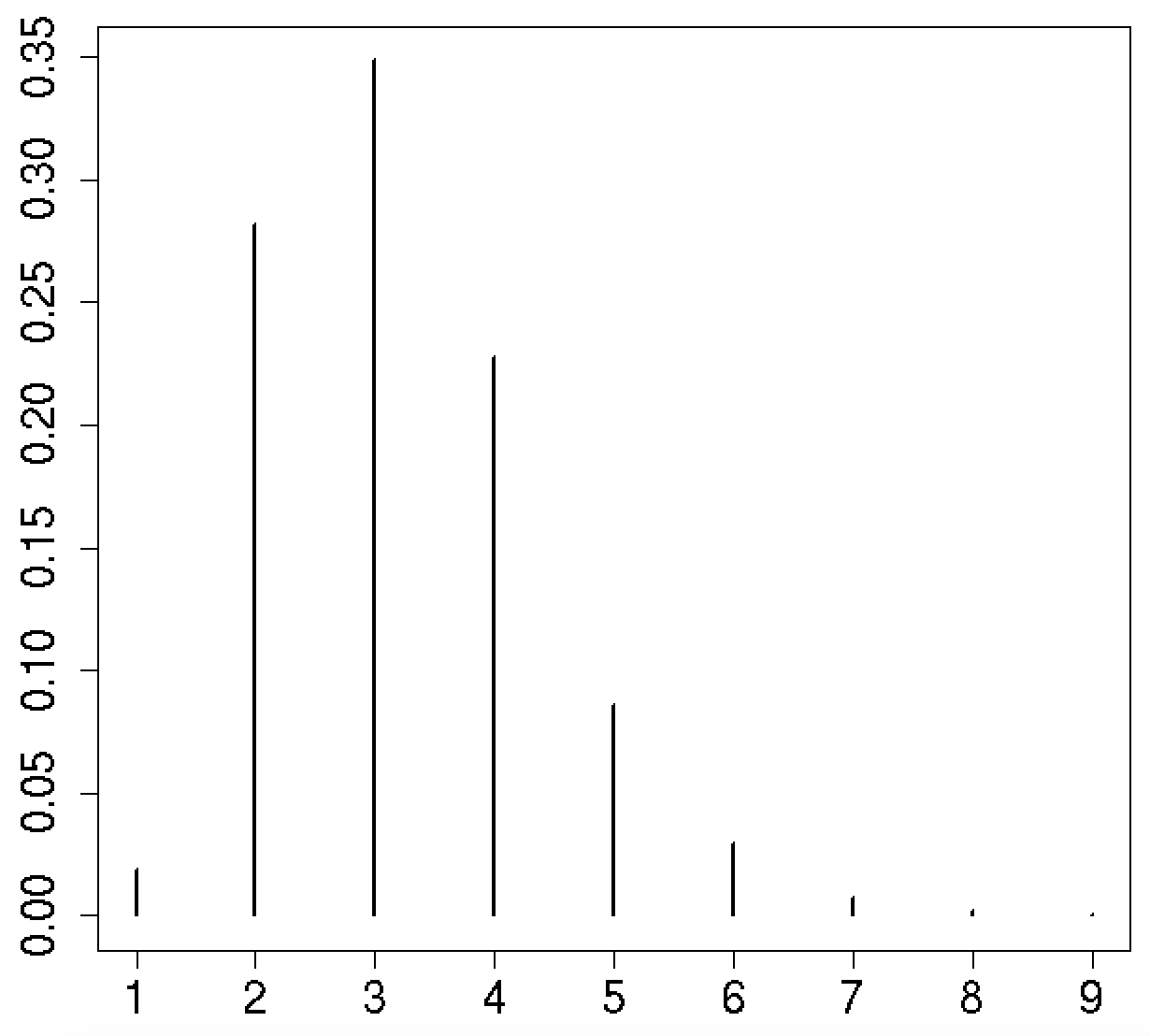}} 
\subfigure[]%
{\includegraphics[width=0.325\textwidth,height=0.35\textwidth]{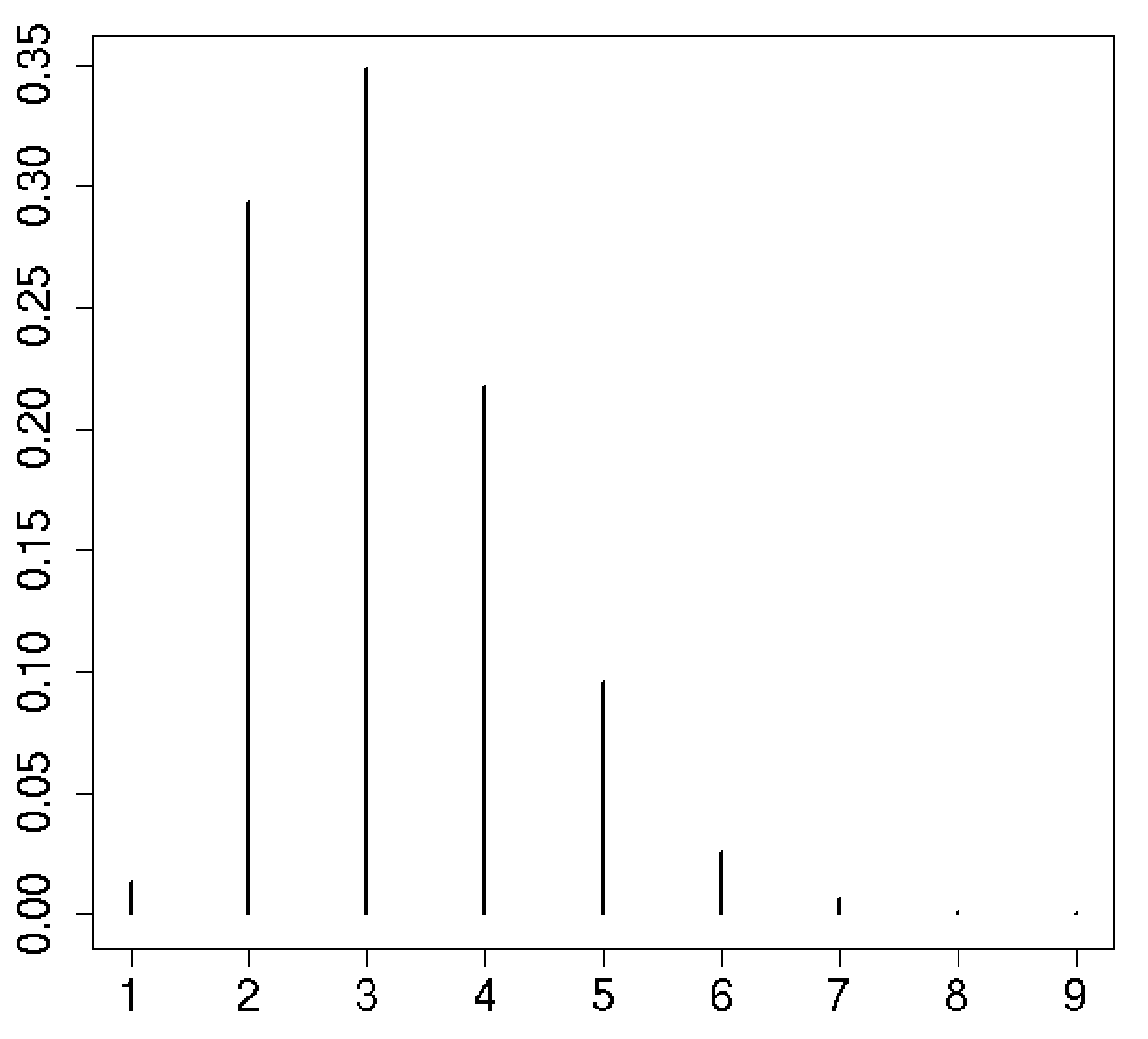}} 
\subfigure[]%
{\includegraphics[width=0.33\textwidth,height=0.35\textwidth]{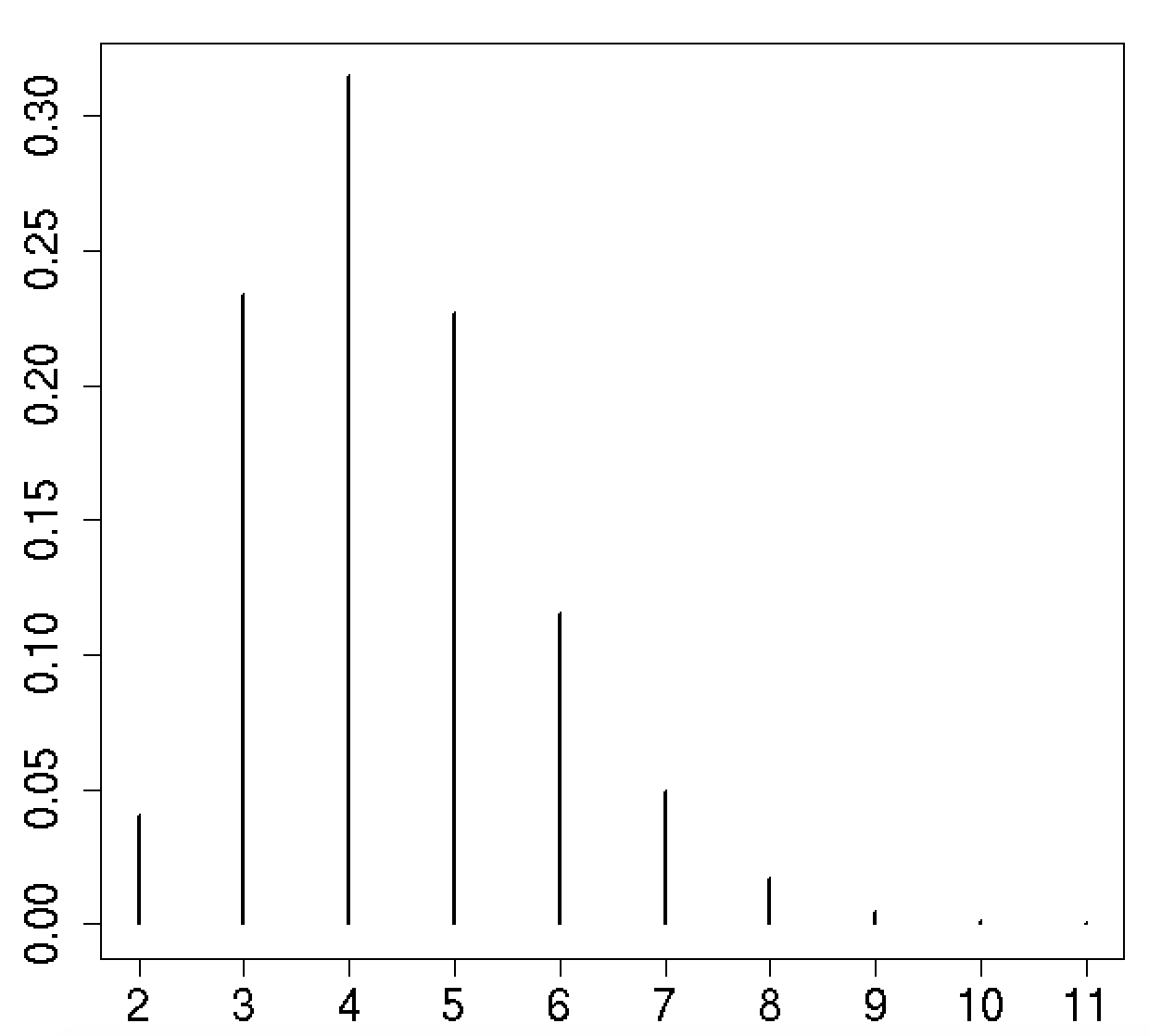}}  
\caption{ \textit{Readmission} dataset: posterior distribution of $K$, with $f(Y_{i1},\ldots$, $Y_{i j-1})= (Y_{i 1}+\cdots+ Y_{i j-1})/(j-1)$ (a) and $f(Y_{i1},\ldots,Y_{i j-1})= \left(Y_{i1}\times \cdots\times Y_{i j-1} \right)^{1/(j-1)}$ (b). Panel c displays the posterior distribution of $K$ using the AR(2) model.}
\label{fig:read_K}
\end{figure}

We compare now the results of the nonparametric AR(2) model for the random effects $\alpha_{ij}$'s as in \eqref{eq:randarp}-\eqref{eq:p-dimG0}, selected in the previous section, with models built using 
different choices of $f$. In particular we consider two summary statistics:  $f(Y_{i1},\ldots, Y_{i j-1})= (Y_{i 1}+\cdots+ Y_{i j-1})/(j-1)$ and $f(Y_{i1},\ldots,Y_{i j-1})= \left(Y_{i1}\times \cdots\times Y_{i j-1} \right)^{1/(j-1)}$. The goal is to understand if higher order temporal dependency can be approximated 
by an AR(1)-like process built on some appropriate function of past observations as described in 
\eqref{eq:rand_eff}.
Figure~\ref{fig:read_K} displays the posterior of $K$, the number of components in the mixture \eqref{eq:randarp}-\eqref{eq:mult_m} under different alternatives. In particular, the three plots show that 
 the posterior modes of $K$ are 2 or 3 with a probability of around 30\% for the AR(1)-type models. 
%but also 4 clusters is a plausible value, 
On the other hand,  Figure~\ref{fig:read_K}(c), referring to the AR(2) model, suggests the existence of 3, 4 or 5 groups.

\begin{figure}[h!]
\centering 
{\includegraphics[width=\textwidth,height=0.6\textwidth]{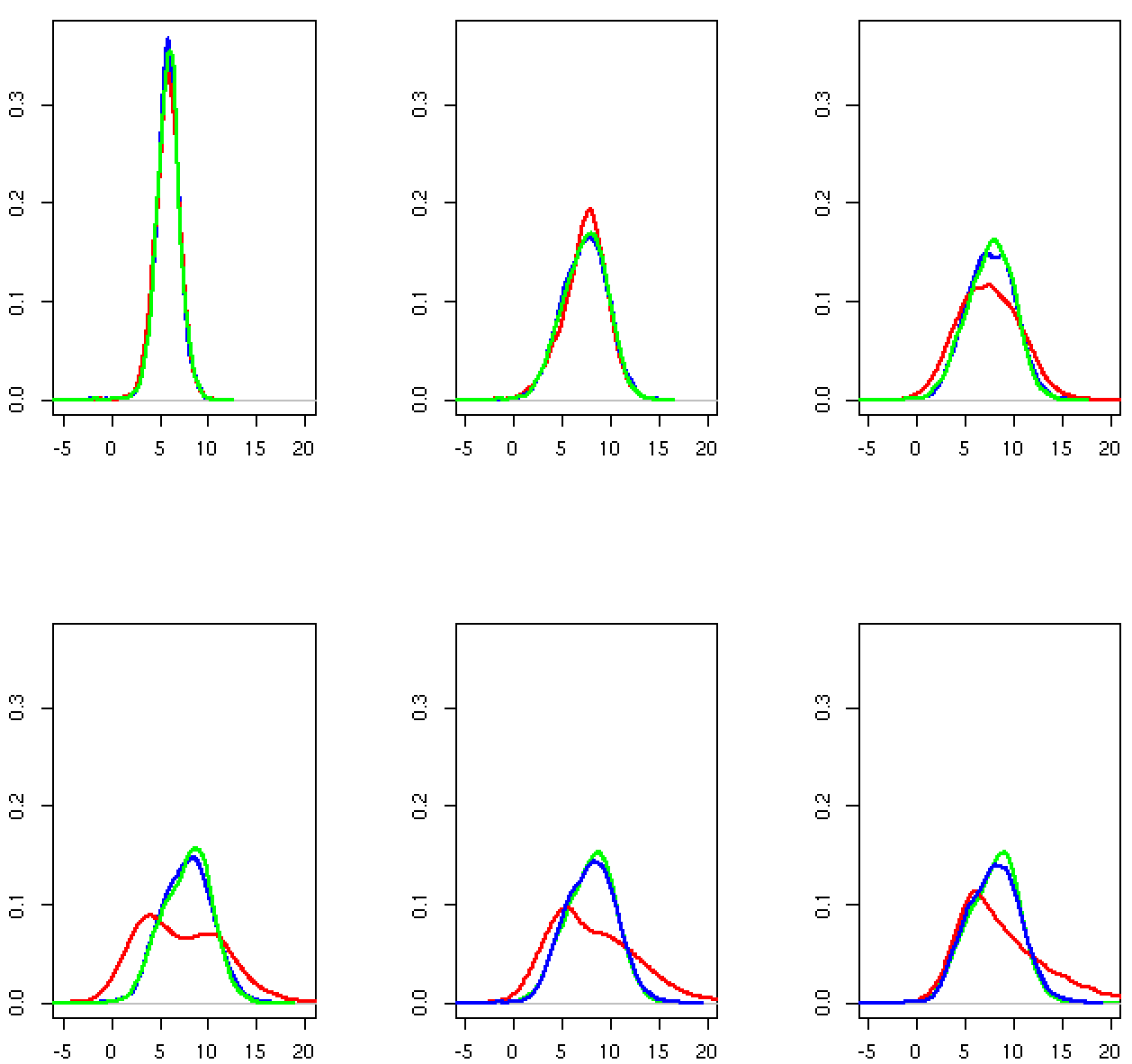}} 
\caption{Readmission dataset: posterior predictive distributions of $Y_{ij}$, $j,\ j=1,\dots,6$.The green and blue lines represent AR(1)-type models, with
$f(Y_{i1},\ldots, Y_{i j-1})= (Y_{i 1}+\cdots+ Y_{i j-1})/(j-1)$ and $f(Y_{i1},\ldots,Y_{i j-1})= \left(Y_{i1}\times \cdots\times Y_{i j-1} \right)^{1/(j-1)}$,   respectively , and the  red distribution indicates AR(2) model.}
\label{fig:read_pred}
\end{figure}

In Figure~\ref{fig:read_pred} we present  posterior predictive distributions of $Y_{ij}$ for a hypothetical  new subject,  for each time $j,\ j=1,\dots,6$, setting the values of the covariates to the empirical mode. 
%} the posterior predictive distributions of $Y_{ij}$ for each time $j,\ j=1,\dots,6$ are displayed. 
%The colors green and blue represent AR(1)-type models, respectively 
%$f(Y_{i1},\ldots, Y_{i j-1})= (Y_{i 1}+\cdots+ Y_{i j-1})/(j-1)$ and $f(Y_{i1},\ldots,Y_{i j-1})= \left(Y_{i1}\times \cdots\times Y_{i j-1} \right)^{1/(j-1)}$, while the color red indicates AR(2) model.
From the figure, it is evident that the two AR(1)-type models produce very similar results.
Obviously, for $j=1$ and $j=2$ the three distribution are almost identical, as the models are closer. For  $j>2$, it is clear that the posterior predictive distributions of $Y_{ij}$ have a larger variance and are more skewed under the AR(2) model. This experiment shows that it is not straightforward to approximate higher order dependency using summary statistics. 

\subsection{Posterior inference on the regression parameters}
We now discuss the inference on the regression parameters in order to understand how covariates influence the recurrent event process.  Although some covariates are fixed and do not vary over time, we still assume that their effect can be different in time and therefore we estimate a different vector of regression coefficient for all covariates in the model for each waiting time $j$, $1,\ldots,6$. Covariates \textit{chemo} and \textit{sex} are binary variables, while \textit{dukes} and \textit{charlson} are 3 levels categorical variables and we need to introduce 2 dummy variables for each of them in the model, with baseline set to A--B for \textit{dukes} and to 0 for \textit{charlson}. Therefore, the final covariate vector for individual $i$ is given by ${\bm x}_i=(x_{i1},x_{i2},x_{i3},x_{i4},x_{i5},x_{i6})=$(indicator for chemotherapy, indicator for female, indicator for  \textit{dukes} equal to C, indicator for  \textit{dukes} equal to D, indicator for \textit{charlson} in 1 --2, indicator for \textit{charlson} = 3). The vector of regression parameters ${\bm \beta}_j=(\beta_{1j},\beta_{2j},\beta_{3j},\beta_{4j},\beta_{5j}, \beta_{6j})  $ for each gap time $j$, $j=1,\ldots,J =6$,  is therefore 6-dimensional.

% ($x_{i1}=1$) and the gender (=1 if woman), while  $(x_{i3}, x_{i4})$ are dummy variables for the ordinal covariate \textit{dukes} (baseline: A-B) and  $(x_{i5}, x_{i6)}$ indicate the  \textit{charlson} covariate (baseline: 0); these last two covariates are time-varying.   

\begin{figure}[!ht]
\centering 
\subfigure[Regression coefficients of \textit{chemo}] %of $\beta_{11},\beta_{12},\beta_{13},\beta_{14},\beta_{15}$]
{\includegraphics[width=0.49\textwidth,height=0.25\textwidth]{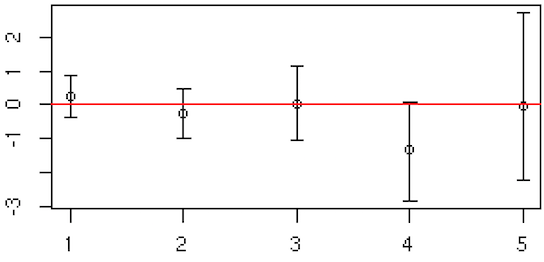}} 
\subfigure[Regression coefficients of \textit{sex}]
{\includegraphics[width=0.49\textwidth,height=0.25\textwidth]{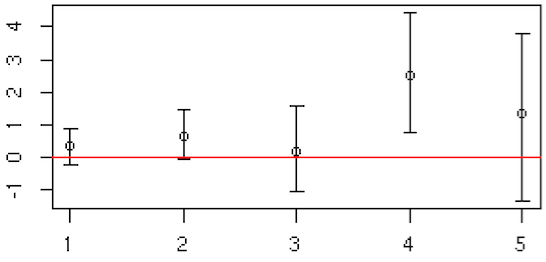}} \\
\subfigure[Regression coefficients of \textit{dukes = C}]
{\includegraphics[width=0.49\textwidth,height=0.25\textwidth]{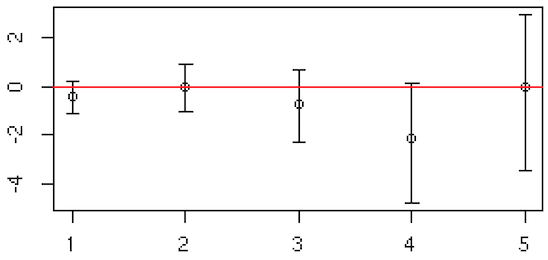}} 
\subfigure[Regression coefficients of \textit{dukes = D}]
{\includegraphics[width=0.49\textwidth,height=0.25\textwidth]{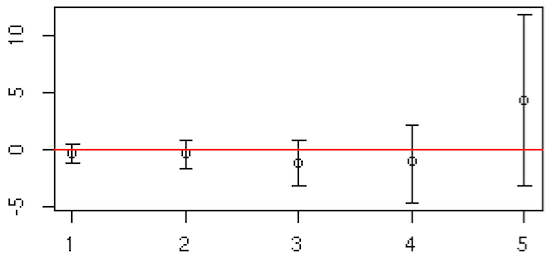}} \\
\subfigure[Regression coefficients of \textit{charlson = 1--2}]
{\includegraphics[width=0.49\textwidth,height=0.25\textwidth]{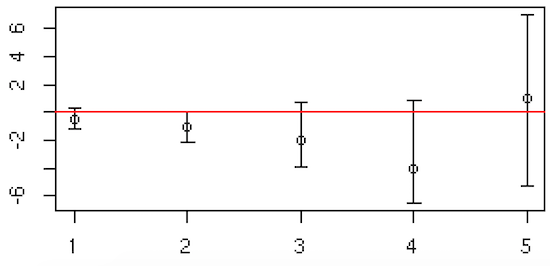}} 
\subfigure[Regression coefficients of \textit{charlson = 3}]
{\includegraphics[width=0.49\textwidth,height=0.25\textwidth]{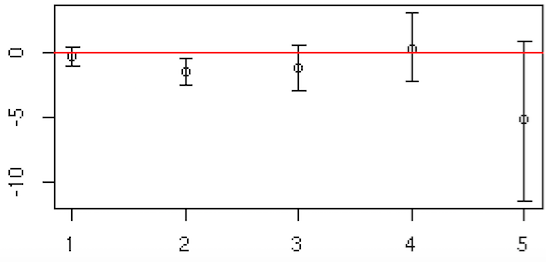}} 
\caption{Posterior 95\% credible interval for the regression parameters of each covariate  across the first five gap times.}
\label{fig:read_CI}
\end{figure}

 %In this section, we denote with  $\tilde{\bm{\beta}}_i$ the vector of the parameter relative to the $i^{th}$ covariate for each gap time. Remark that, on the other hand, we denoted with $\bm{\beta}_i$ the vector of all the covariate parameters for the $i^{th}$ gap time.
Figure~\ref{fig:read_CI} shows the 95\% credible intervals for the posterior marginals of the regression parameters; in particular, each panel displays the posterior CIs of the regression parameter of each covariate for the first 5 gap times, i.e. of $\beta_{r1},\beta_{r2},\beta_{r3},\beta_{r4},\beta_{r5}$, where $r$ denotes the covariates. 
For example, 
Figure~\ref{fig:read_CI}(a) shows that there is no evident effect of chemotherapy on any  gap time. However,  the CI of  $\beta_{14}$ is concentrated on negative values, which means that chemotherapy reduces the fourth waiting time between hospitalisations.
%, i.e. the time elapsed between the fourth and the fifth recurrent events.% In general, however, there does not seem to be an effect of chemotherapy on the outcome.
% We can deduce the following considerations:
%\begin{itemize}
%\item $\tilde{\bm{\beta}}_1,$ which captures the effect of the chemotherapy on the gap times, does not seem to be significant for the first gap times. However, at the fourth gap time the CI for  $\beta_{14}$ is concentrated on negative values, which means that chemotherapy reduces the fourth waiting time between hospitalisations, i.e. the time elapsed between the fourth and the fifth recurrent events. In general, however, there does not seem to be an effect of chemotherapy on the outcome.
%\item $\tilde{\bm{\beta}}_2,$ which measures the effect of sex on the gap times,
%Figure~\ref{fig:read_CI}(b) indicates that women have mainly larger waiting times. This is more evident for the second and the fourth component, but a trend is visible at each recurrent event. 
%\item $\tilde{\bm{\beta}}_3,$ the first dummy variable relative to 
%The Dukes stage of the tumour %(stage C versus the baseline stage A-B), 
i%s never significantly different from 0; see Figure~\ref{fig:read_CI}(c)-(d).  
%\item $\tilde{\bm{\beta}}_5,$  
%\textcolor{red}{SISTEMARE meglio questa parte di commenti qui sotto }\\
%the first dummy variable relative to the Charlson Index of the patient (index 1-2 versus the baseline index 0), has negative medians, apart from the last gap time. Therefore patients with index 1-2 will experience more frequent recurrent events with respect to the ones with index 0.
%\item $\tilde{\bm{\beta}}_6,$ 
%the second dummy variable relative to the Charlson Index of the patient (index 3 versus the baseline index 0), is mostly negative. Therefore patients with index 3 will have shorter gap times with respect to the ones with index 0.
In general, credible intervals are larger for the last gap times, as expected, since few individuals have a large number of  events. The regression coefficients at time $j=6$ are not shown as the credible intervals are not comparable with those of the previous times.  
%\end{itemize}

\section{Urinary Tract Infection dataset}
\label{sec:realdata_UTI} 
\begin{figure}[h!]
\centering 
%\subfigure[]%
{\includegraphics[width=0.49\textwidth,height=0.35\textwidth]{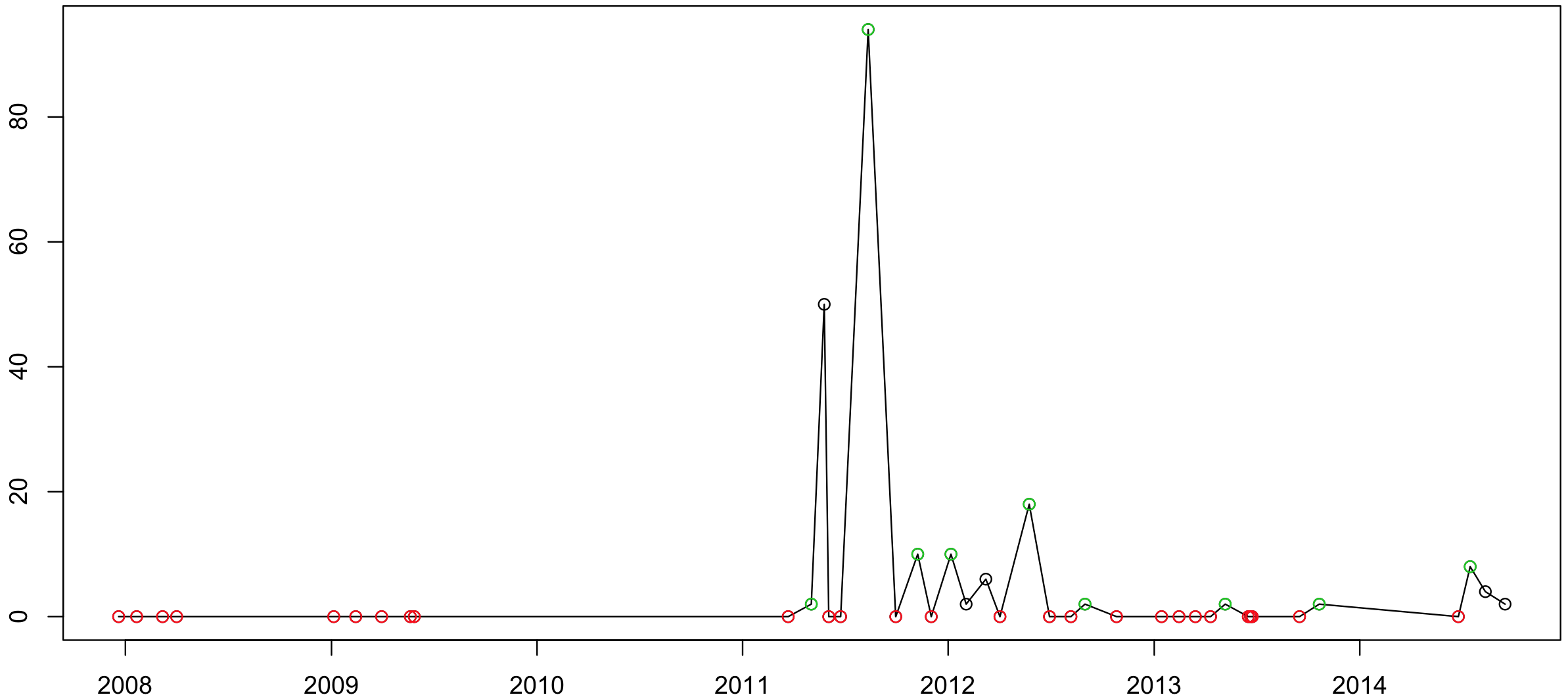}} 
%\subfigure[]%
{\includegraphics[width=0.49\textwidth,height=0.35\textwidth]{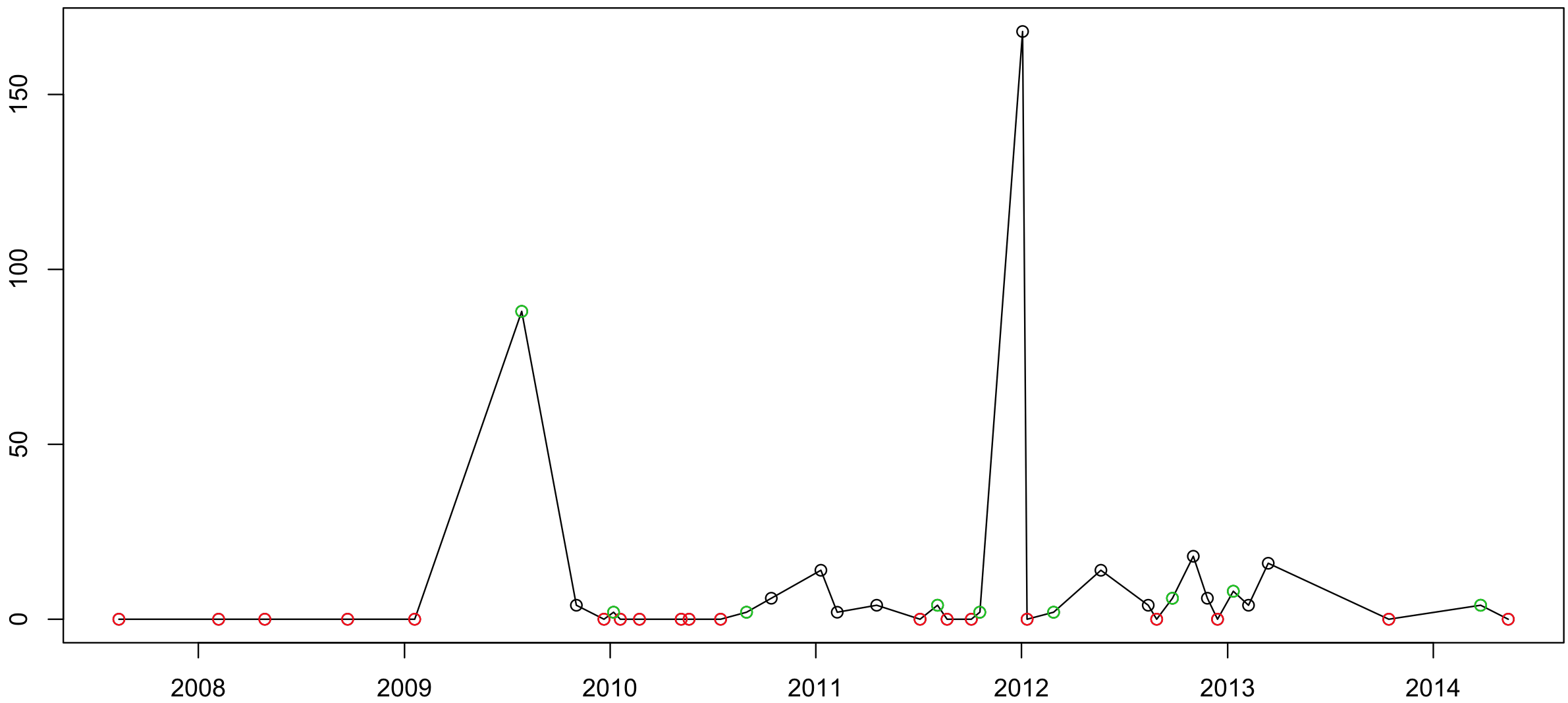}} 
\caption{Recurrent events for two patients: the last waiting time of the patient on the left is observed,  while that of the patient on the right is censored. Red circles denote zero WBC at the visit while green circles denote WBC greater than 0.}
\label{fig:wait_UTI_paz2_3}
\end{figure}

We consider data on patients at risk of urinary tract infection (UTI). 
The best clinical marker  of UTI available is pyuria, i.e. White Blood Cell count (WBC) $\mu l^{-1}$ $\geq 1$, detected by microscopy of a fresh unspun, unstained specimen of urine (\cite{khasriyaetal2010, kupelianetal2013}). 
Let $T_{i0}$ correspond to the first visit attendance at the \textit{Lower Urinary Tract Service Clinic}  (Whittington Hospital, London, UK)  and let $T_{ij}$ be the time of the $j-th$ new infection for the patient $i$. Note that at time 0, all patients suffer of UTI.
For each patient and at each visit the result of the microanalysis of a sample of urine has been recorded in terms of the WBC count. Presence of WBC in the urine (regardless of the quantity) indicates the presence of Urinary Tract Infection. 
We include in the analysis only female patients with at least two waiting times, giving a total of 
$N=306$  patients. The number of observations with exactly $j$ gap times is displayed in Table~\ref{tab:num_UTI}.
\begin{table}[!ht]
 \centering
\begin{tabular}{|c|c|c|c|c|c|c|c|c||c|}
\hline
 j & 2 & 3 & 4 & 5 & 6 & 7 & 8 & 9&TOT\\
\hline
$n_j$  & 121 & 89 & 54 & 21 & 10 & 6 & 2 & 3 & 306\\
\hline
\end{tabular}
\caption{Number of observations for $j$ gap times, $j = 1, \dots, 9$.} 
\label{tab:num_UTI}
\end{table}
\medskip
We note that 85 subjects out of 306 are right-censored with respect to their last gap time. Since the proportion of censored data is considerable, we have taken censoring into account and modified the likelihood appropriately.
Figure~\ref{fig:wait_UTI_paz2_3} displays the recurrent events of two randomly selected patients, in which the last waiting time of the patient in the left panel is observed,  while that of the patient in  the right panel is censored. Indeed, the first patient suffers of  infection at her last visit, while the second  patient has a WBC counts equal to zero implying that a new infection will happen necessarily  after  her last visit.

We fit model  \eqref{eq:meanY}, including for each patient a 5-dimensional vector of time-varying covariates:  %\eqref{eq:rand_eff}
%\eqref{eq:DP2}, where, in this case 
%\begin{eqnarray*}
%{\bm x}_{ij}^T {\bm\beta}_j = x_{ij1}^T {\beta_{j1}}+x_{ij2}^T {\beta_{j2}}+x_{ij3}^T {\beta_{j3}}+x_{ij4}^T {\beta_{j4}}+x_{ij5}^T {\beta_{j5}}
%\end{eqnarray*}
a continuous covariates representing 
the standardized age of the patient $i$ during gap time $j$ and  four   binary variables denoting  the presence, during the $j$-th gap time, of 
urgency, pain, stress incontinence and voiding symptoms (=1 if the symptom is present, 0 otherwise).  Therefore, the final covariate vector for individual $i$ is given by ${\bm x}_i=(x_{i1},x_{i2},x_{i3},x_{i4},x_{i5})=$(\textit{age}, indicator for urgency, indicator for incontinence, indicator for pain, indicator for voiding).
%The components of each vector ${\bm x}_{j}$ are the following:
%\begin{itemize}
%\item $x_{ij1}$ indicates the standardized age of the patient $i$ during gap time $j$;
%\item $x_{ij2}$ is binary, equal to 1 if the patient $i$ presents the urgency symptom during gap time $j$;
%\item $x_{ij3}$ is  binary, equal to 1 if the patient $i$ presents the pain symptom during gap time $j$;
%\item $x_{ij4}$ is binary, equal to 1 if the patient $i$ presents the stress incontinence symptom during gap time $j$;
%\item $x_{ij5}$ is binary, equal to 1 if the patient $i$ presents the voiding symptom during gap time $j$;
%\end{itemize}
%The covariate vector ${\bm x}_{ij}$ of patient $i$ at time $j$ is 5-dimensional, with all the components evolving in time. 
Descriptive statistics of the covariates are given in Table~\ref{tab:UTI}.
\begin{table}[!ht]
\centering
\begin{tabular}{|c|c|c|}
\hline 
Covariate & Mean & Standard Deviation\\ 
\hline 
age & 53.87 & 16.01\\
\hline
presence of urgency symptoms &0.56  &0.50 \\
\hline 
presence of incontinence symptoms &0.21  &0.41 \\
\hline 
presence of pain symptoms &0.47  &0.50 \\
\hline 
presence of voiding symptoms &0.45  &0.50 \\
\hline 
\end{tabular}
\caption{Descriptive statistics of the covariates of the UTI dataset.} 
\label{tab:UTI}
\end{table}

In the analysis we set the  prior hyperparameters in \eqref{eq:priorAR1} %\eqref{eq:priorbeta}-\eqref{eq:G0} 
 in order to specify vague prior distributions:
\begin{eqnarray*}
\beta_0^2 &=& 1000\\
\sigma &\sim&  \Uc(0,10)\\
\tau &\sim&  \Uc(0,10)\\
%M &=& 1\\
\sigma_g^2 &=& 10, \quad a_Z = 3, \quad b_Z = 3.
\\ M&=&1 \ . 
\end{eqnarray*}

\subsection{Posterior Inference}

\begin{figure}[h!]
\centering 
\subfigure[]%
{\includegraphics[width=0.40\textwidth,height=0.35\textwidth]{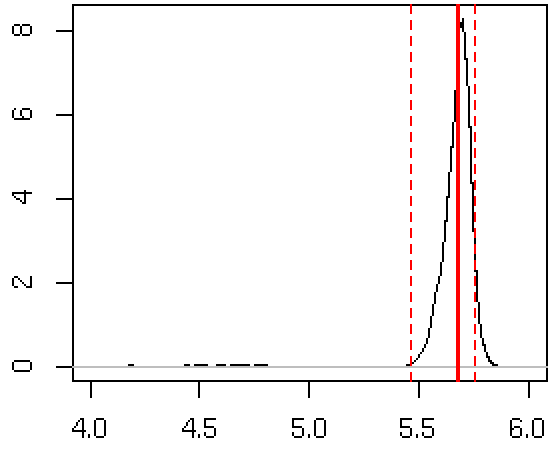}} 
\subfigure[]%
{\includegraphics[width=0.40\textwidth,height=0.35\textwidth]{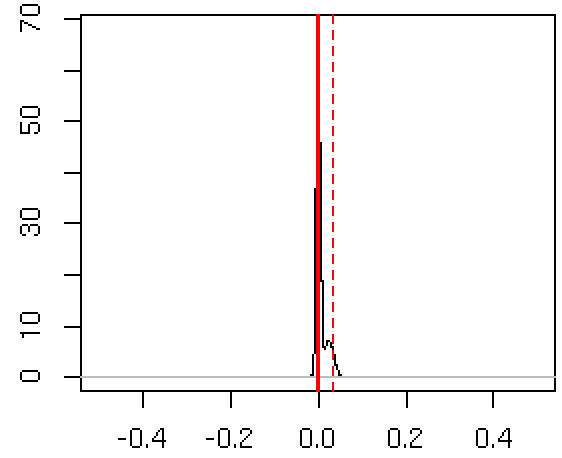}} 
\subfigure[]%
{\includegraphics[width=0.40\textwidth,height=0.35\textwidth]{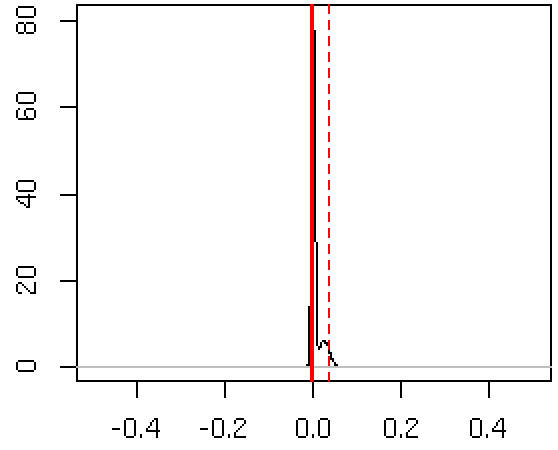}}  
\subfigure[]%
{\includegraphics[width=0.40\textwidth,height=0.35\textwidth]{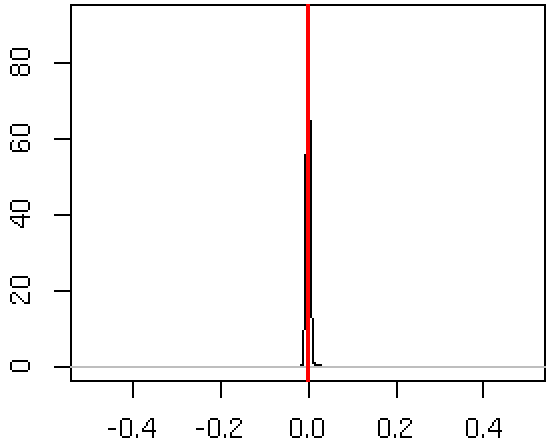}} 
\caption{\textit{UTI} dataset: predictive marginal distributions of  $m_{i0}$(a), $m_{i1}$(b), $m_{i2}$(c) and $m_{i3}$(d).  Dashed vertical lines denote 0.05 and 0.95 posterior quantiles, while the bold vertical line is the posterior median.}
\label{fig:real_SSVS_m}
\end{figure}

We run the model for the three choices of function $f$ described in Subsection 2.1. We obtain 
similar  posterior predictive marginal distributions for $m_{i0}$ and $m_{i1}$, as well as the  same posterior inference for  $K$. %(Con i tre modelli AR(1) abbiamo ottenuto le stesse posterior). 
In particular,  a posteriori, the marginal distribution of  $m_{i1}$ is concentrated around 0, indicating independence between   between gap times. 
This result is confirmed also by performing inference on the order of dependence using both approaches introduced in   
Section~\ref{sec:3}.
The posterior predictive marginal distributions of $m_{i,l}$, for $l=0,1,2,3$, obtained with spike and slab variable selection, is displayed in Figures~\ref{fig:real_SSVS_m}: panel $(b)$, $(c)$ and $(d)$ show that  the 
posterior predictive marginal distributions of $m_{i,1}$, $m_{i,2}$, $m_{i,3}$ are all  concentrated around $0$. In addition, also specifying directly a  prior on $p$ with $P=3$ leads to  a posterior distribution for the order of temporal dependence with  mode in 0 (result not shown).

%\section{Conclusion and further work}
\section{Conclusion}
\label{sec:concl}
In this work we have proposed novel Bayesian nonparametric approaches for modelling  waiting times between recurrent events. 
Time-dependency is taken into account through  the specification of an autoregressive model on the random effects governing the distributions of the gap times. To allow for clustering of patients, overdispersion and outliers, we introduce Dirichlet process mixtures as random effects distribution. Covariates may be easily included in this framework.

The strategy we adopt is flexible and allows testing  for the order of dependence among random effect at different times, that is a key feature of the nonparametric AR(p) model.
%nd then the choice of $f$ in Eq.~\eqref{eq:rand_eff}. 
%We illustrate the model through an application to recurrent hospitalizations of cancer patients.
We propose two different methods to test the order of dependence: spike and slab variable selection and direct prior on the order of dependence. In the first case we can simply modify the base measure of the DP, whereas with the second technique, we elicit a prior on the order $p$ of the autoregressive process and then, conditioning on $p$, we set a Dirichlet Process prior of appropriate dimension for the parameters of the AR(p) model.
%After that, our aim was analyze the posterior marginal distribution of the latent variables of DP and the posterior of the number of clusters, i.e. the number of groups that the latent variables done.
%Both in simulations and in real dataset they have produce the same results.

We can introduce the time-dependency in different ways.  The simplest and probably most natural way consists of assuming that  the random effects at time $j-1,\ldots,j-p$ influence the behaviour of the random effect at time $j$.  We then investigate the possibility of approximating higher order of dependency using summary statistics of past observations. Our results show that the choice of summary statistics is crucial and not obvious and that the approximation worsens as the number of gap times increases. As such, this topic  will of object of future research, possibly borrowing ideas from the Approximate Bayesian Computation literature.

%We investigated the performance of our model with some simulations and then, we illustrated the model in real-data applications.

This type of model strategy can be extended to other fields of application; in particular it is  straightforward 
to adapt the proposed approach to model multiple time series analysis  \citep[see][]{NieQui16JTS, DiLucca_etal13}. In fact, in this case, the data consist in  $N$ time
series  $\mathbf{Y}_{i} =(Y_{i1}, \ldots, Y_{in_i})$, where $i$ denotes the time series and $n_i$ is the number of observations for each series. The likelihood for each time series can be expressed as in 
\eqref{eq:meanY} and temporal dependence may be introduced as in  \eqref{eq:rand_eff}-\eqref{eq:DP2} with appropriate choice of the function $f(\cdot)$.  Moreover, the proposed model can also be used as building block in a hierarchy to describe the relationship between recurrent events and survival up to a terminating event, for example in a survival regression context. This latter extension is object of on-going investigation.

%Time window is not necessarily the same for each subject in the sample.  Here we will model the random dynamic multiple time series as a collection of dependent univariate series.

\bibliography{Reference_Marta}
\bibliographystyle{ba}
\end{document}